\documentclass[a4paper,fleqn]{cas-sc}

\usepackage[authoryear]{natbib}
\usepackage{url,hyperref,lineno,microtype,subcaption}
\usepackage[onehalfspacing]{setspace}
\usepackage{graphicx}

\usepackage{indentfirst}
\usepackage{array,multirow}
\usepackage{fancyhdr}
\usepackage{float}
\usepackage{booktabs}
\usepackage{multirow}
\usepackage{here}
\usepackage{comment}
\usepackage{siunitx}
\usepackage{algorithmicx,algorithm,algpseudocode}
\usepackage{amsmath,amsfonts,bm}
\usepackage{fontawesome5}
\usepackage{caption}
\usepackage{subcaption}
\usepackage{xcolor}
\usepackage{pifont}
\usepackage[figuresright]{rotating}
\renewcommand{\thefootnote}{\roman{footnote}} 
\usepackage[font=small,labelfont=bf]{caption}

\newcommand{\etal}{et~al.\ }
\newcommand{\eg}{e.\,g.,\ }
\newcommand{\ie}{i.\,e.,\ }

\hypersetup{
    colorlinks=false,
    linkcolor=blue,
    filecolor=magenta,      
    urlcolor=cyan,
    pdfpagemode=FullScreen
    }

\def\tsc#1{\csdef{#1}{\textsc{\lowercase{#1}}\xspace}}
\tsc{WGM}
\tsc{QE}
\tsc{EP}
\tsc{PMS}
\tsc{BEC}
\tsc{DE}

\begin{document}
\let\WriteBookmarks\relax
\def\floatpagepagefraction{1}
\def\textpagefraction{.001}

\shorttitle{Is Silent eHMI Enough for APMV?}

\shortauthors{Liu et~al.}

\title [mode = title]{Is Silent eHMI Enough? A Passenger-Centric Study on Effective eHMI for Autonomous Personal Mobility Vehicles in the Field
}                      
\author[a]{Hailong Liu}[orcid=0000-0003-2195-3380]
\cormark[1]
\cortext[cor1]{Corresponding author.~~~~Email: liu.hailong@is.naist.jp (H.Liu)}

\author[b]{Yang Li}[orcid=0000-0001-7100-8040]
\author[c]{Zhe Zeng}[orcid=0000-0001-9188-2181]
\author[d]{Hao Cheng}[orcid=0000-0002-3254-4796]
\author[e]{Chen Peng}[orcid=0000-0001-7285-3123]
\author[a]{Takahiro Wada}[orcid=0000-0002-4518-8903]

\address[a]{Graduate School of Science and Technology, Nara Institute of Science and Technology, 8916-5 Takayama-cho, Ikoma, Nara, 630-0192, Japan}

\address[b]{Institute of Human and Industrial Engineering, Karlsruhe Institute of Technology, Engler-Bunte-Ring 4,Karlsruhe,76133,Germany}

\address[c]{Dept. Human Factors, Institute of Psychology and Education, Ulm University, Albert-Einstein-Allee 45, 89081, Ulm, Germany}

\address[d]{Faculty of Geo-Information Science and Earth Observation, University of Twente, 7500 AE Enschede, The Netherlands}

\address[e]{Institute for Transport Studies, University of Leeds, 36-40 University Rd, Leeds LS2 9JT, United Kingdom.}

\begin{abstract}
Autonomous Personal Mobility Vehicle~(APMV) is a miniaturized autonomous vehicle designed to provide short-distance mobility to everyone in pedestrian-rich environments.
By the characteristic of the open design, passengers on the APMV are exposed to the communication between the eHMI deployed on APMVs and pedestrians.
Therefore, to ensure an optimal passenger experience, eHMI designs for APMVs must consider the potential impact of APMV-pedestrian communications on passengers' subjective feelings.
To this end, this study discussed three external human-machine interface (eHMI) designs, \ie 1) graphical user interface (GUI)-based eHMI with text message (eHMI-T), 2) multimodal user interface (MUI)-based eHMI with neutral voice (eHMI-NV), and 3) MUI-based eHMI with affective voice (eHMI-AV), from the perspective of APMV passengers in the communication between APMV and pedestrians.
In the riding field experiment (N=24), we found that eHMI-T may be less suitable for APMVs.
This conclusion was drawn based on passengers' feedback, as they expressed an awkward feeling during the ``silent time'' when the eHMI-T provided information only to pedestrians but not to passengers.
Additionally, these two MUI-based eHMIs with voice cues had their own advantages, \ie eHMI-NV has an advantage in pragmatic quality, while eHMI-AV has an advantage in hedonic quality.
The study also highlights the necessity of considering passengers' personalities when designing eHMI for APMVs to enhance their experience.
\end{abstract}


\maketitle

\section{INTRODUCTION}

\subsection{Autonomous Personal Mobility Vehicle}
Autonomous Personal Mobility Vehicle~(APMV) is a miniaturized vehicle with automation functions ranging from the SAE levels 3 to 5 automated driving systems that can decide their own driving behavior without passenger involvement.
Currently, most well-known APMVs are indeed developed based on electric wheelchairs~\footnote{\textit{WHILL Autonomous} developed by WHILL Inc.: \url{https://youtu.be/vJWhwNnUPRs}} or semi-open small vehicles~\footnote{\textit{RakuRo} developed by ZMP Inc.: \url{https://youtu.be/lWhbJ0rBwjM}}. 
It is important to note that these APMVs are not only developed for the elderly people or people with disabilities; they can be used by anyone for short-distance mobility.

As shown in Fig.~\ref{fig:image}, unlike autonomous cars, APMVs can be widely used in shared spaces with other traffic participants, such areas include sidewalks, shopping centers, stations, school campuses, and other mixed transportation areas~\citep{Yoshinori2013, ali2019smart,liu2020_what_timeing,liu2022implicit}, to facilitate the mobility of passengers.
As a result, APMVs will inevitably interact with more vulnerable road users like pedestrians without any protection.

\subsection{Interaction Issues of APMV in Shared Spaces}

In the aforementioned pedestrian-rich shared spaces, pedestrians may have a low perceived safety when an APMV uses implicit communication, such as changes in maneuvers, to interact with them, as they may misunderstand the APMV's driving intentions~\citep{liu2020_what_timeing,liu2022implicit}.
This confusion can lead to undesirable consequences (\eg lower acceptance levels), or even dangers (\eg crashes).
Therefore, it is crucial to explore effective ways of communication between APMVs and pedestrians, in order to accurately convey their intentions to each other and ensure safe and efficient interactions between them.
One solution to improve APMV-pedestrian communications is to equip an external Human-Machine Interface (eHMI) on the APMV. 

\subsection{External Human-Machine Interface and Its Challenge in User Experience for APMV Passengers}

Based on the literature review, we found that current eHMI designs and research primarily focus on the interactions between autonomous vehicles (AVs) and pedestrians.
These eHMIs often rely on the visual cues through light bars, icons, text, and ground projections~\citep{bazilinskyy2021should, liu2021importance, dey2021towards, li2021autonomous, li2023av}.
As mentioned in the review article~\citep{brill2023external}, auditory-based eHMI has been under-studied within the eHMI literature when compared to vision-based eHMIs.
A few works comparing the effects of vision-based and auditory-based eHMIs of AVs on pedestrians' user experience~\citet{dou2021evaluation} and reactions to warning messages~\citet{ahn2021comparative}, as well as \citet{haimerl2022evaluation} explored the feasibility of auditory eHMI of AVs for pedestrians with intellectual disabilities.
Moreover, \citet{kreissig2023blinking} assess the impact of vision-based and auditory-base eHMIs on the perception of surrounding pedestrians during the autonomous parking of a driver-less E-Cargo bike.
To the best of our knowledge, there is presently no research addressing the design of eHMI for APMVs and investigating the user experiences of vision-based and auditory-based eHMI for APMVs.

\begin{figure}[t]
  \centering
  \includegraphics[width=0.75\linewidth,trim=0 0 0 160,clip]{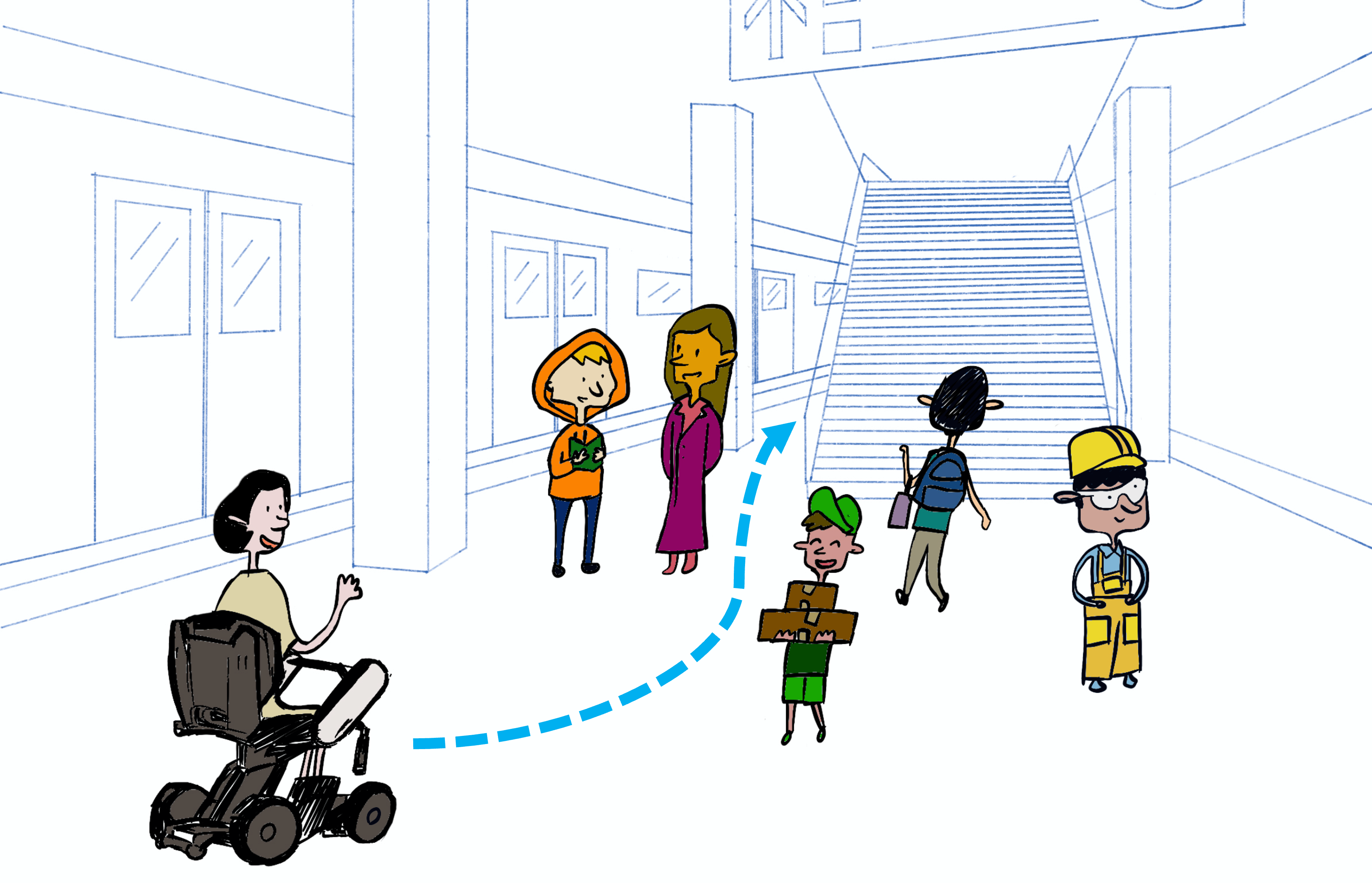}
  \vspace{-4mm}
  \caption{An APMV encounters pedestrians in a shared space.}
  \label{fig:image}
  \vspace{-4mm}
\end{figure}

To apply eHMI designs from AVs to APMVs, we must first carefully consider the distinct characteristics that set AVs and APMVs apart.
As shown in Fig.~\ref{fig:image}, unlike AV--pedestrian interactions, APMVs have unique characteristics in their interactions with pedestrians:
i) APMVs are often in closer proximity to pedestrians, given their much smaller sizes and lower driving speeds than AVs, especially in indoor areas such as shopping centers and airports, where pedestrians are populated.
ii) In addition to close proximity, passengers on the APMV are exposed to the communication between the eHMI deployed on APMVs and pedestrians by the characteristic of the open design.
Therefore, to ensure an optimal passenger experience, eHMI designs for APMVs must consider the potential impact of APMV-pedestrian communications on passengers' subjective feelings.

Based on the characteristics mentioned above, a few studies have used eHMI to improve the interactions between pedestrians and APMVs.
For example, Watanabe~\etal used a projector to project the trajectory of an APMV's movement onto the ground, in order to communicate the motion intentions of the APMV to both pedestrians and passengers~\citep{watanabe2015communicating}.
\citet{zhang2022understanding} conducted a series of interviews with wheelchair users, and revealed the existence of two distinct interaction loops, \ie APMV-passenger and APMV-pedestrian.
Moreover, they suggested a range of design concepts, derived from a virtual workshop, for those two interactions, \ie projecting information, such as planned path, on the ground for APMV-passenger interaction, and showing trajectory on a display and using vibrotactile feedback for APMV-passenger interactions.
Furthermore, based on the related works above, we can clearly see that the eHMIs of APMVs could impact the experience of both passengers and pedestrians. 
However, most of the current eHMI designs for APMVs only aim to improve pedestrian experience, passengers' experience has not yet received much attention or been widely discussed.

\subsection{Purpose and Research Questions}

To address the research gaps mentioned in the previous section, this paper explores communication designs in the interactions between APMVs and pedestrians from the perspective of APMV passengers.
As the APMV is for single-passenger use, which is highly associated with the passenger's own preference, we further consider the passenger's personality for the eHMI designs.

In addition to commonly used graphical user interface (GUI)-based eHMI designs, voice user interface (VUI) can be a useful addition to the GUI-based eHMI~\citep{sodnik2008user}.
Furthermore, considering the perspective of APMV passengers, compared to only GUI-based eHMIs, the voice cues from multimodal user interface (MUI)-based eHMIs can provide confirmative feedback to passengers about the APMV's communications with pedestrians, avoiding feelings of ignorance.
The voice cues can also serve as an alternative communication channel to broadcast the APMV's intentions to other road users in the vicinity who may be visually occupied.

Therefore, we explored multimodal user interface (MUI)-based eHMIs that incorporate both GUI and verbal message-based VUI on APMVs to communicate with pedestrians.
To the best of our knowledge, we are the first to conduct a field study investigating user experience of APMV passengers using MUI-based eHMIs with voice cues.

To sum up, in this paper, we aim to answer the following three questions through a field study:
\begin{description}
\itemsep0em
\item[RQ 1:] To what extent does the silent GUI-based eHMI for AVs apply to APMVs in terms of passenger's user experience?
\item[RQ 2:] To what extent does the voice cues of MUI-based eHMI apply to APMVs in terms of passenger's user experience?
\item[RQ 3:] Does APMV's eHMI design need to fit the passenger's personality?
\end{description}

\section{METHOD}

We conducted a passenger-centric experiment using a robotic wheelchair as the APMV. 
The purpose of the experiment was to investigate the impact of different visual and auditory eHMIs on passengers' subjective feelings when the APMV communicates with pedestrians.
This study has been carried out with the approval of the Research Ethics Committee of Nara Institute of Science and Technology (NAIST) [No.~2022-I-55].

\subsection{Autonomous Personal Mobility Vehicle (APMV)} \label{sec:APMV}

A robotic wheelchair {\it WHILL Model CR} with an autonomous driving system was used as the APMV.
As shown in Table~\ref{Tab:eHMIs_all}, the APMV was equipped with a LiDAR (Velodyne VLP-16) and a controlling PC allowing it to drive autonomously on a pre-designed route.
In this experiment, its maximum speed was limited to 1+m/s for safety considerations.

\subsection{The eHMI for Visualizing APMV's Driving Behaviors}\label{sec:eHMI_designs_DB}

\begin{figure}[t]
\centering
\captionof{table}{The basic design of APMV's eHMI including LED lights and a display is used to show its driving behaviors.}

\label{Tab:eHMIs_all}
\setlength\tabcolsep{1pt}
\renewcommand{\arraystretch}{0.75}
\centering
\begin{tabular}{@{}ccccc@{}}
\toprule
 Moving  & Decelerating & Stopping & Turning left & Turning right    \\ \midrule
\includegraphics[width=0.18\linewidth]{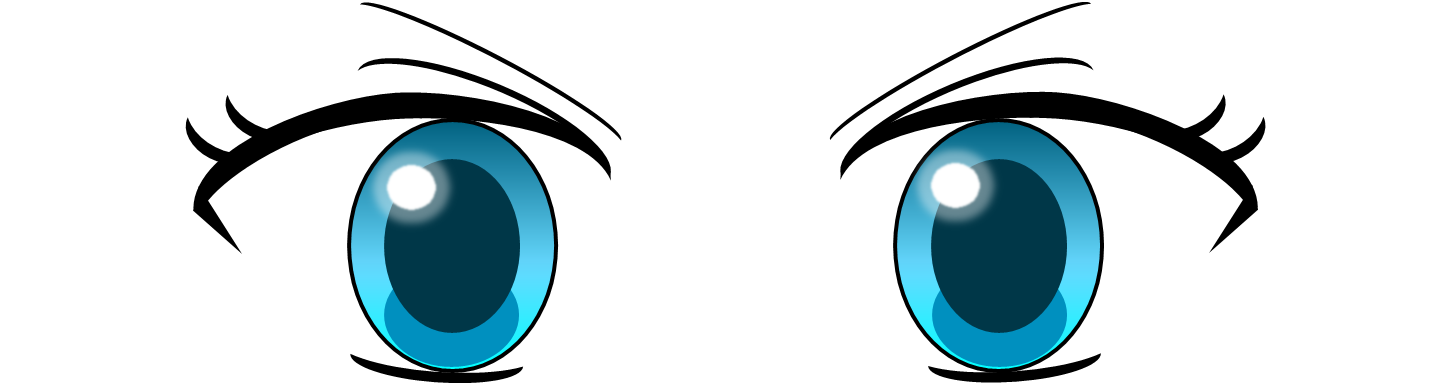}  & \includegraphics[width=0.18\linewidth]{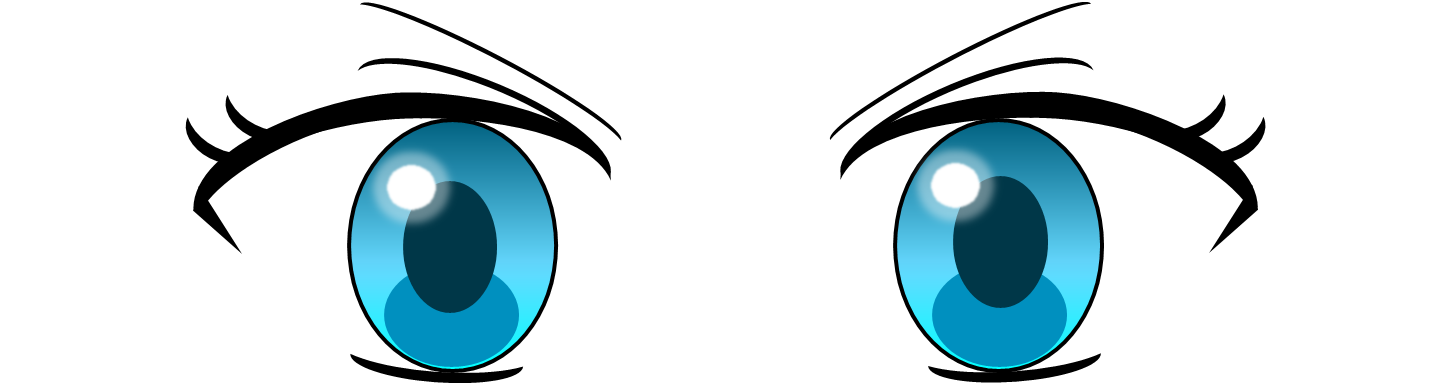}     
&   \includegraphics[width=0.18\linewidth]{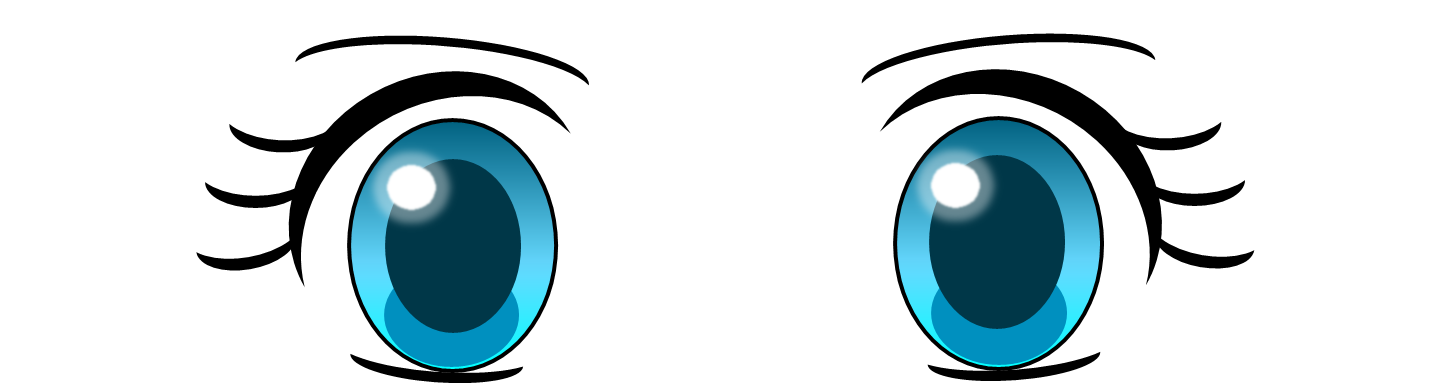}           &  \includegraphics[width=0.18\linewidth]{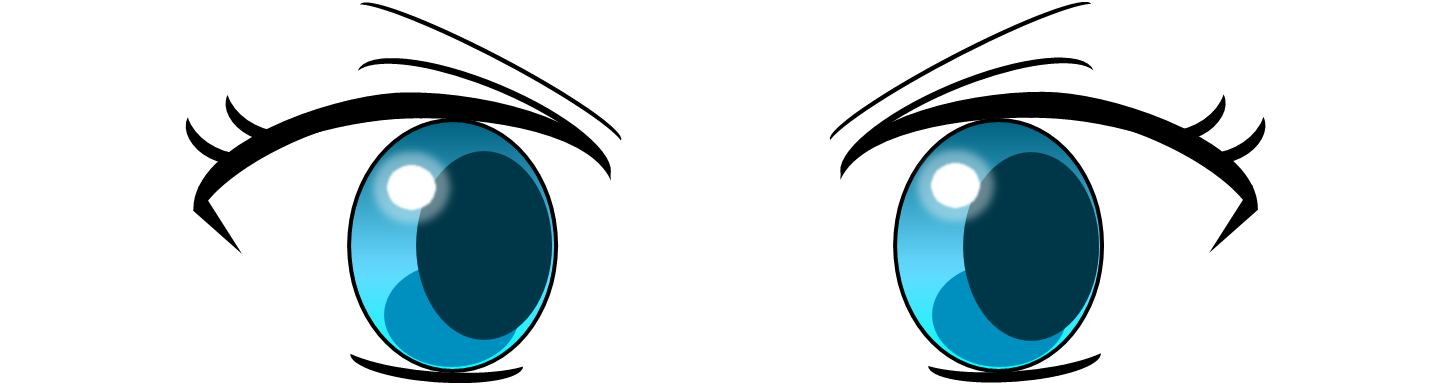}         & \includegraphics[width=0.18\linewidth]{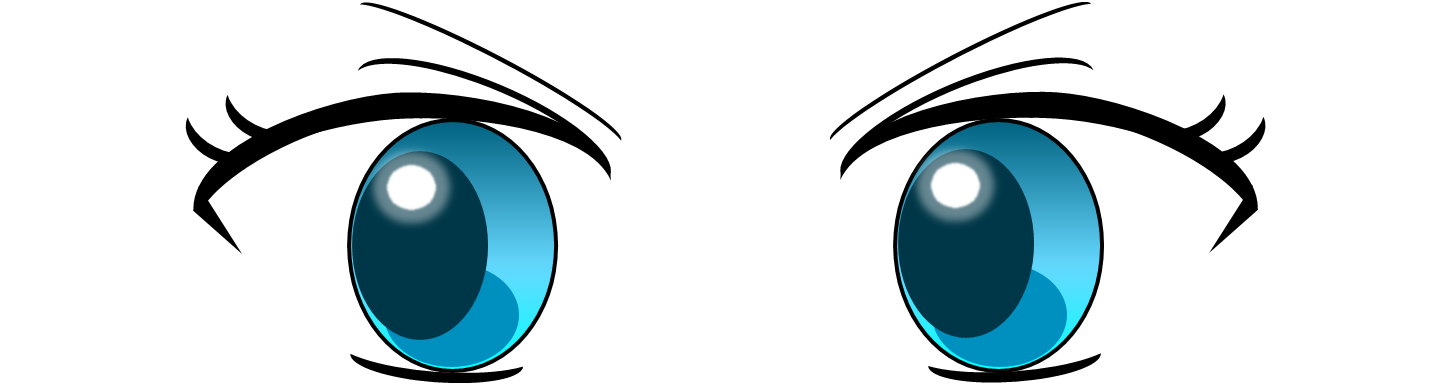}        \\  
\includegraphics[width=0.19\linewidth]{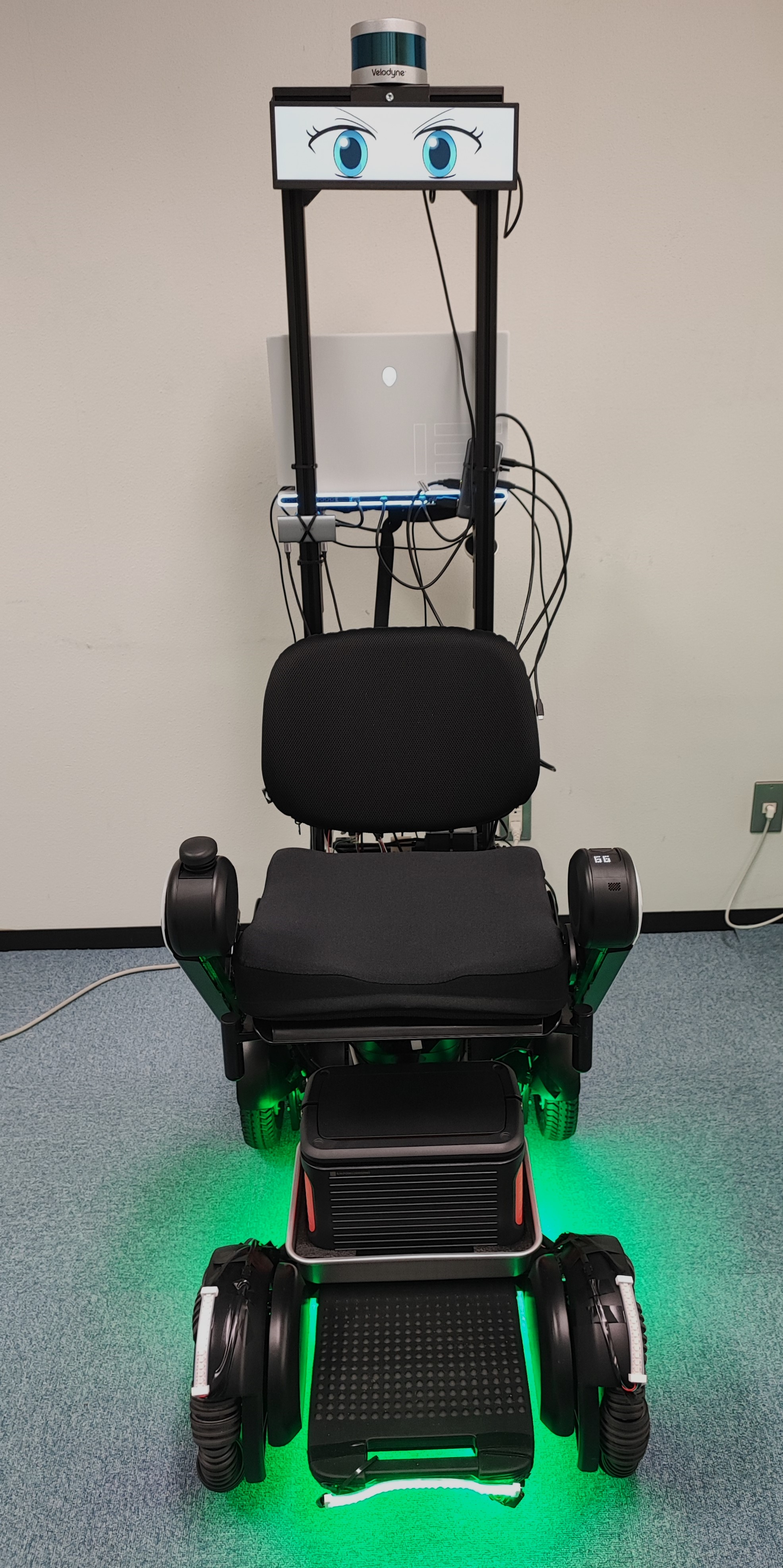}  
& \includegraphics[width=0.19\linewidth]{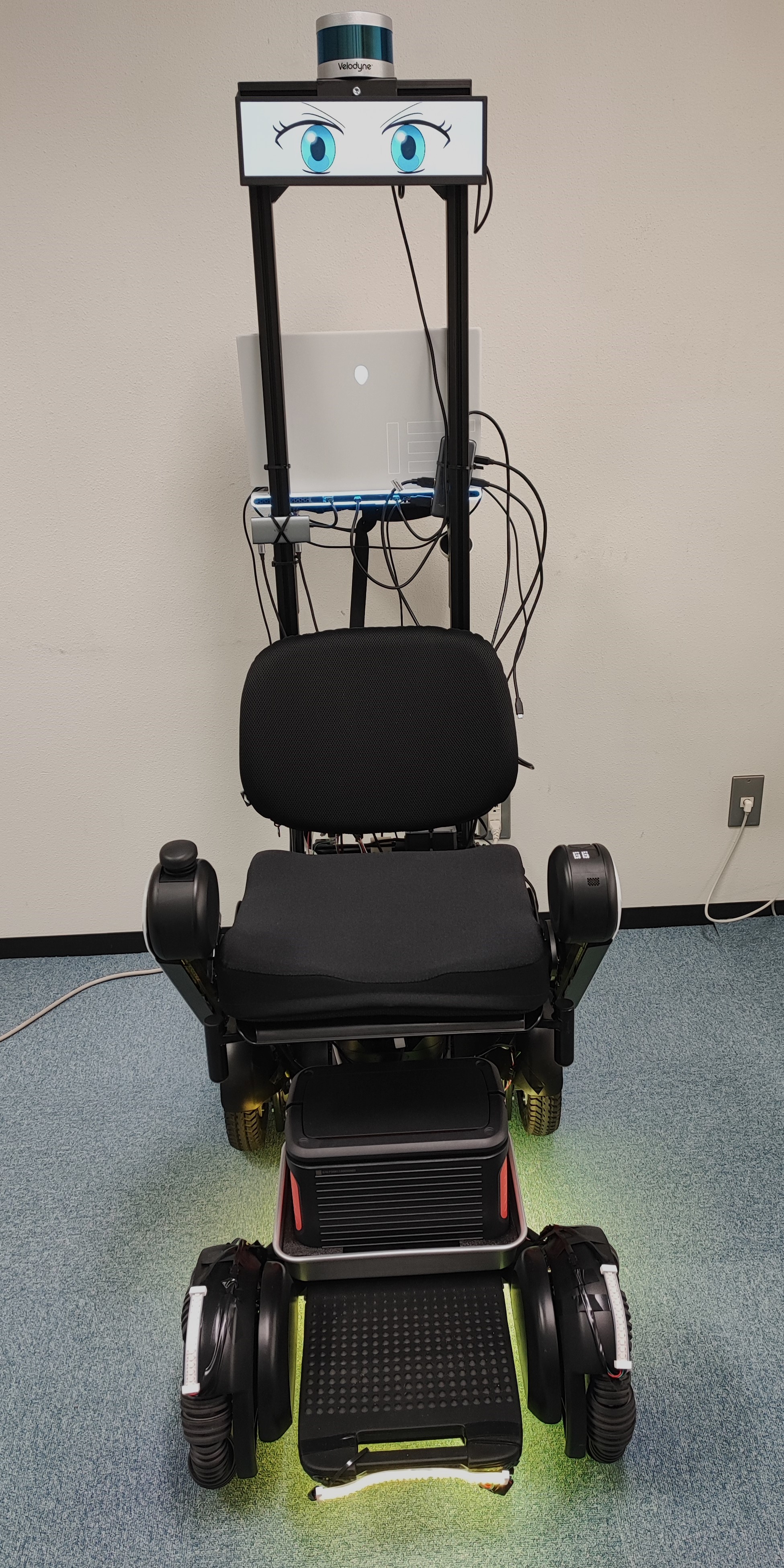}            
& \includegraphics[width=0.19\linewidth]{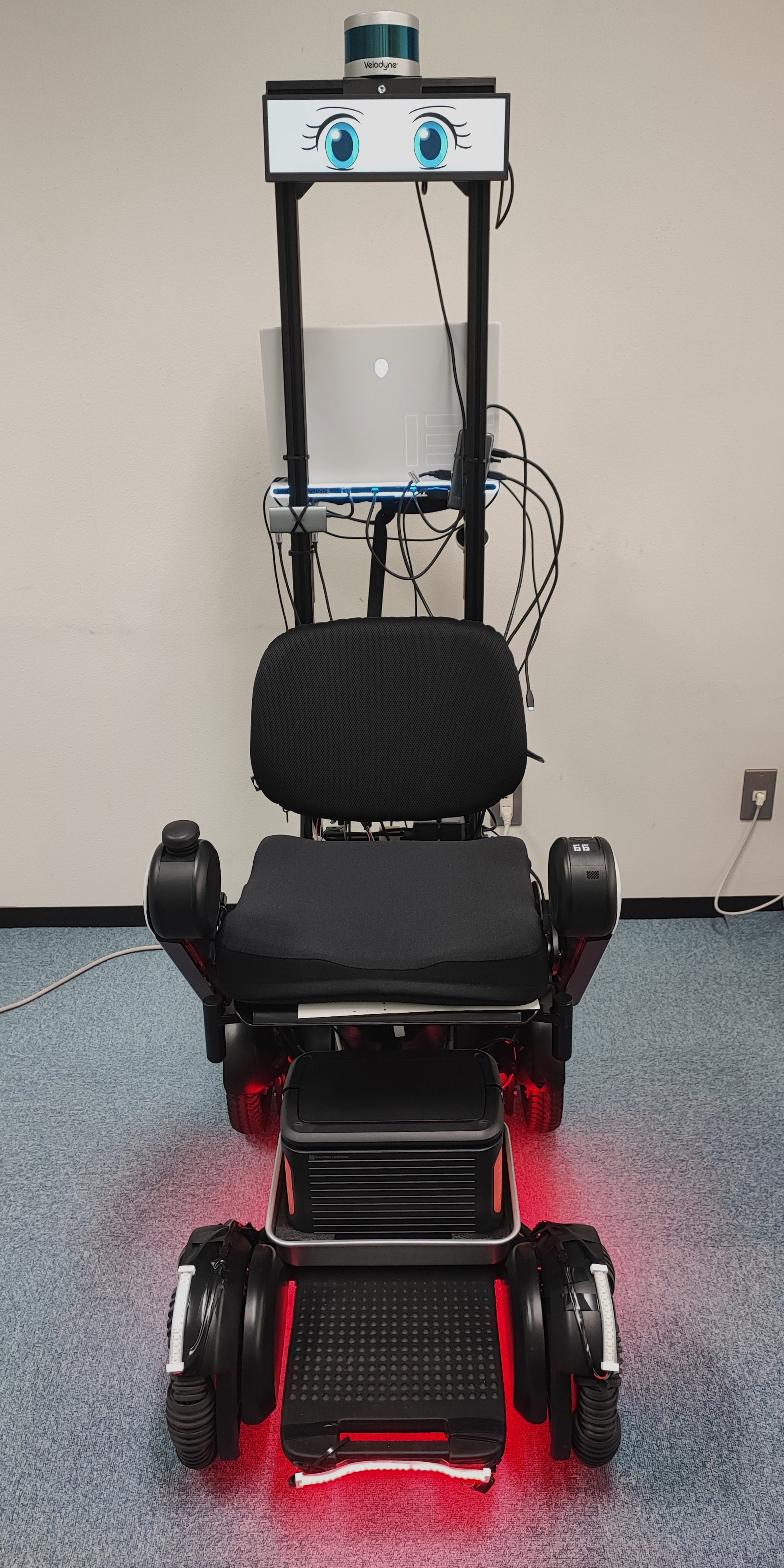} 
& \includegraphics[width=0.19\linewidth]{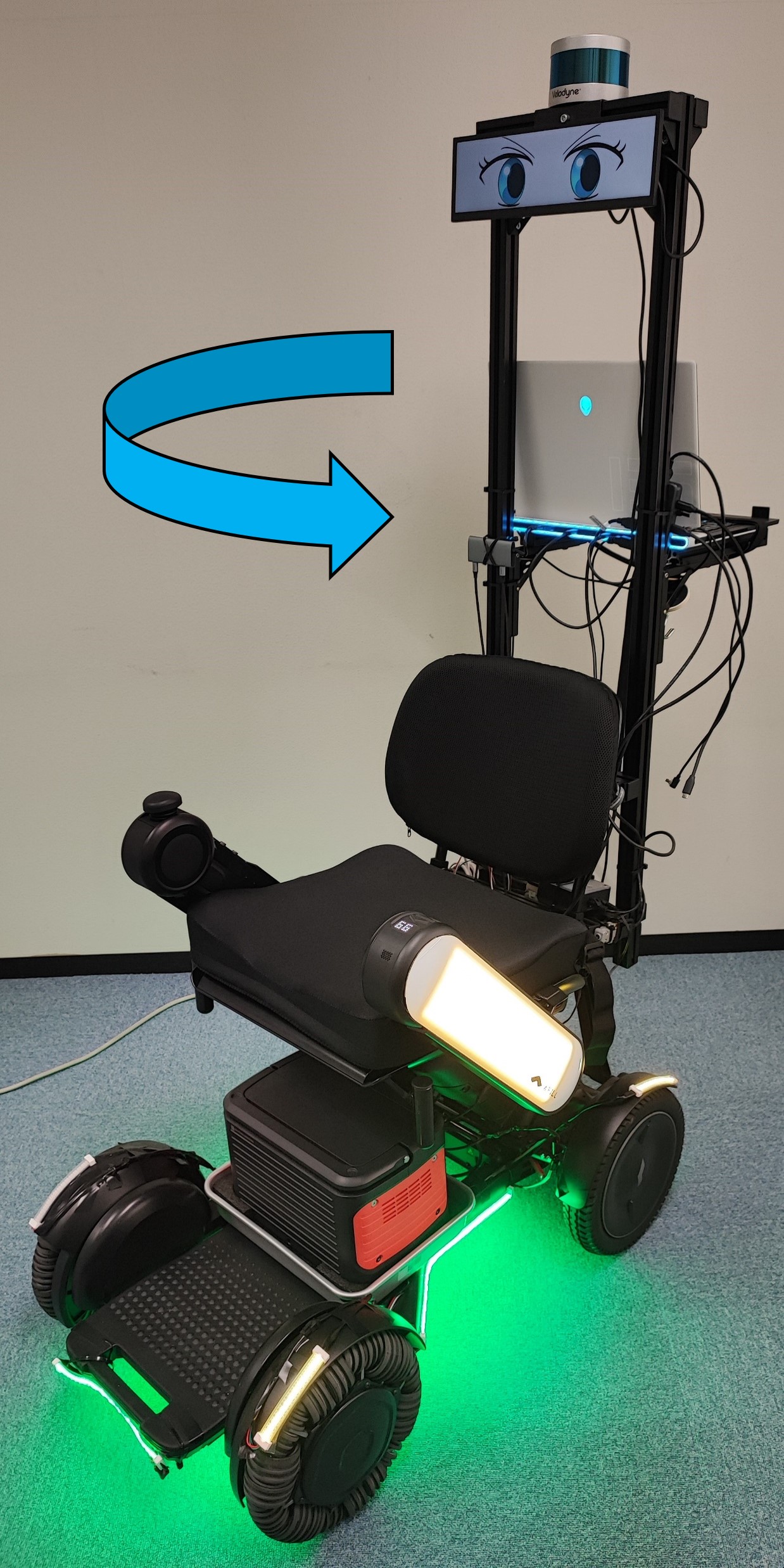} 
& \includegraphics[width=0.19\linewidth]{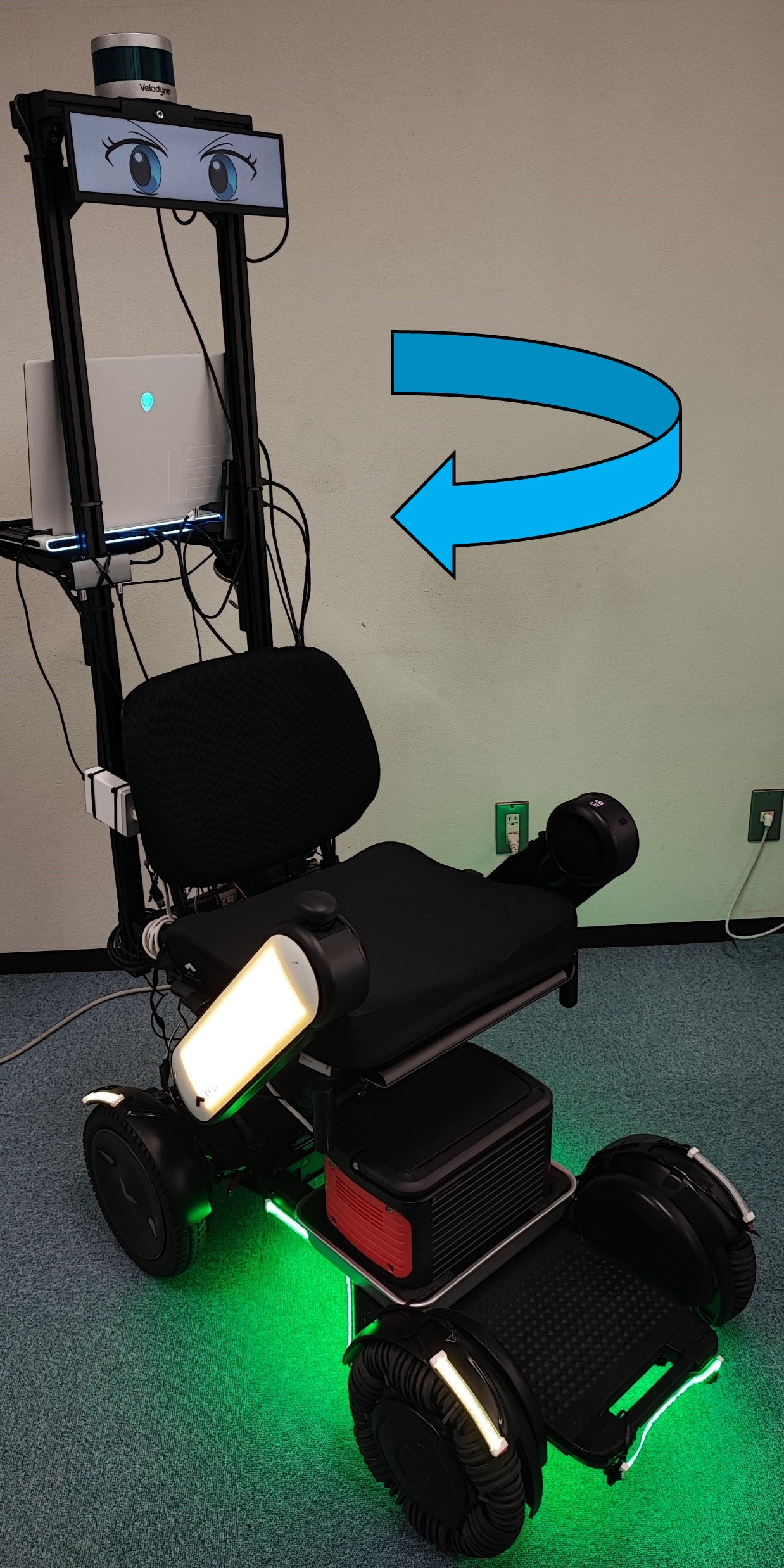} \\ \bottomrule
\end{tabular}
\vspace{-4mm}
\end{figure}

As shown in Table~\ref{Tab:eHMIs_all}, the APMV has a GUI to show the vehicle's driving behaviors via two parts: LED lights on the sides and chassis, and a display above the bodywork.

According to the eHMI design concept in study~\citep{li2021autonomous}, the chassis is equipped with LED lights that show green to move, yellow to decelerate, and red to stop by projecting the respective colored lights onto the ground.
In addition, the yellow LED lights on both sides of the APMV's body and wheels are used as turn signals.

In addition, cartoonish eyes are shown in the display to mimic the observational behaviors of a human driver driving a vehicle.
The eye color is designed to be cyan, showing the autonomous driving mode is ON. 
In particular, when turning, the eyes look in the corresponding direction.
Also, the eyes show a relaxing gaze when stopping, a serious gaze when moving forward, and a concentrated gaze when slowing down.

\subsection{Three eHMIs for Communicating with Pedestrians}
\label{sec:eHMI_designs_C}

\begin{figure}[h!t]
\renewcommand{\arraystretch}{0.7}
\centering
\captionof{table}{Three types of eHMI designs to communicate with pedestrians.}
\label{Tab:eHMIs}
\begin{tabular}{@{}ccccc@{}}
\toprule
 &  & eHMI-T & eHMI-NV & eHMI-AV \\ \midrule
\multirow{2}{*}{Yielding} & GUI &\includegraphics[width=0.21\linewidth]{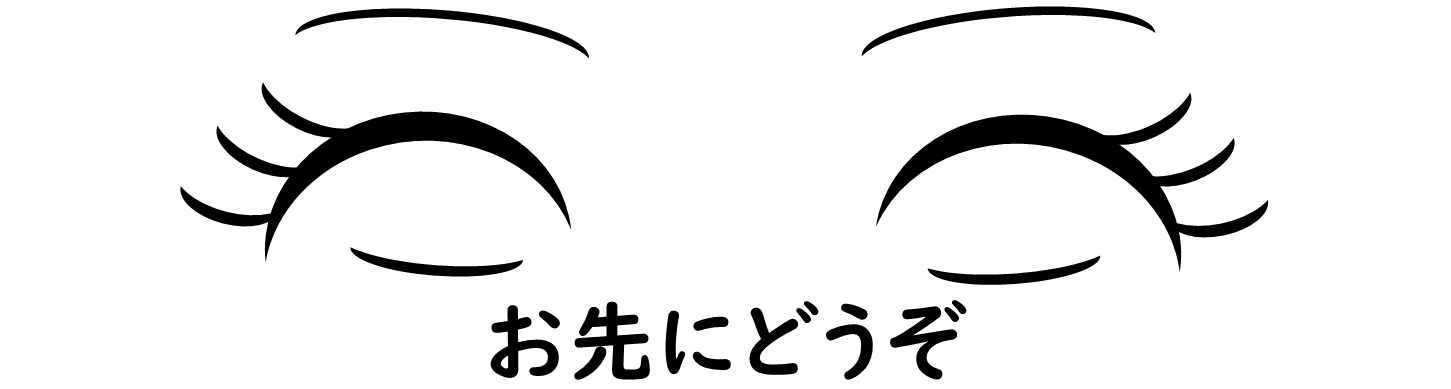}  & \includegraphics[width=0.19\linewidth]{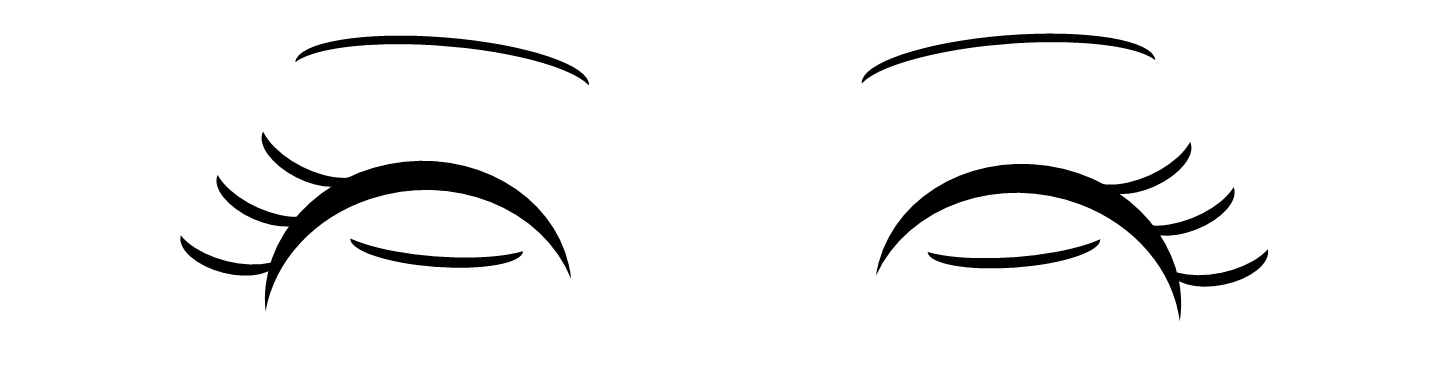}  &\includegraphics[width=0.19\linewidth]{./Fig/eHMI/yield_human.png}  \\ \cmidrule(l){2-5} 
 & VUI & None & \begin{tabular}[c]{@{}c@{}}``After you.''\\ by a neutral voice~\textsuperscript{\ref {Nanami}}
    \end{tabular} & \begin{tabular}[c]{@{}c@{}}``Oh please, after you!''\\ by an affective voice~\textsuperscript{\ref {Nanami}}
   \end{tabular} \\ \midrule
\multirow{2}{*}{Thanks} & GUI & \includegraphics[width=0.21\linewidth]{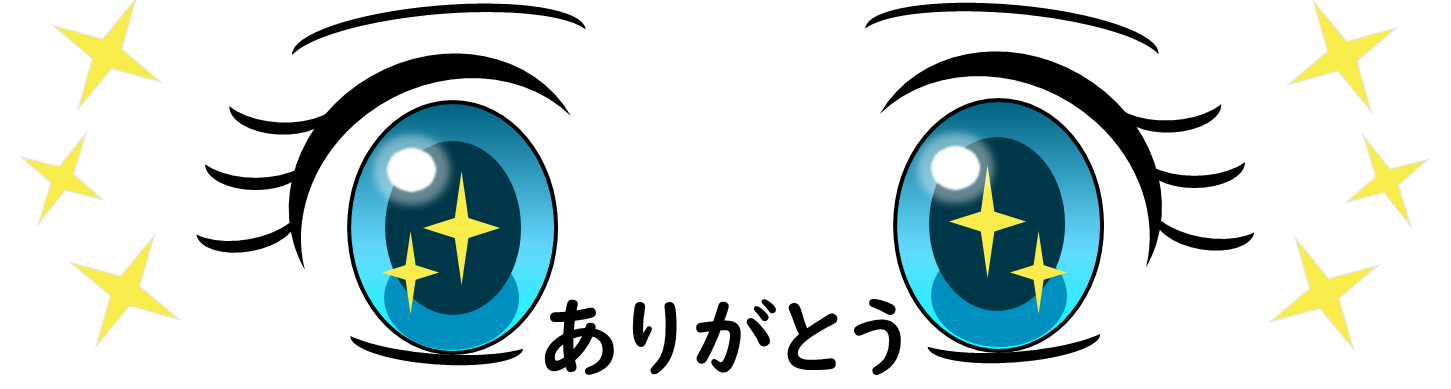} &   \includegraphics[width=0.19\linewidth]{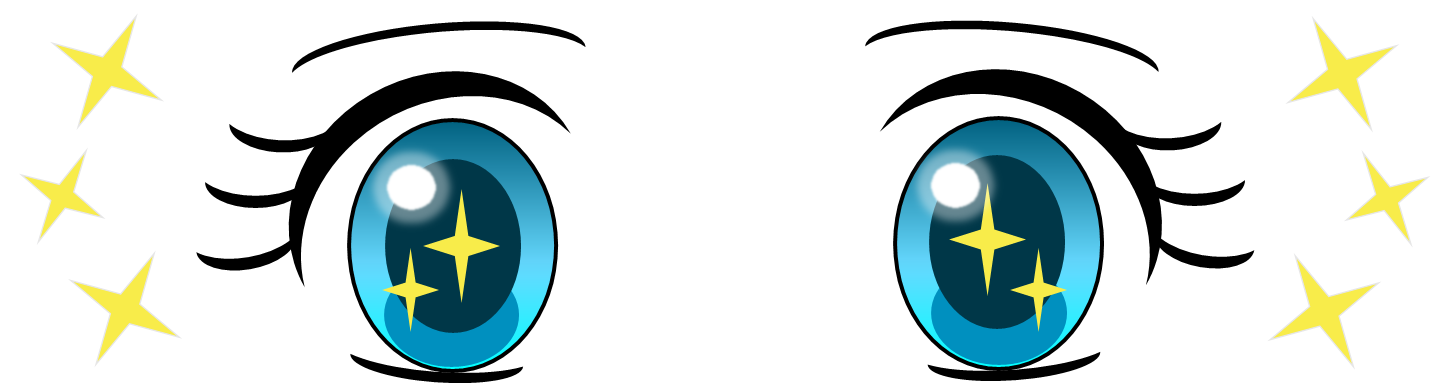}&\includegraphics[width=0.19\linewidth]{./Fig/eHMI/thanks_human.png}  \\ \cmidrule(l){2-5} 
 & VUI & None  &  \begin{tabular}[c]{@{}c@{}}``Thanks.''\\ by a neutral voice~\textsuperscript{\ref {Nanami}}
 \end{tabular} &   \begin{tabular}[c]{@{}c@{}}``You are so kind. Thank you so much!''\\ by an affective voice~\textsuperscript{\ref {Nanami}}
  \end{tabular} \\ \bottomrule
\end{tabular}
\vspace{-3mm}
\end{figure}

In this study, one GUI-based eHMI and two MUI-based eHMIs are designed for the APMV to communicate with a pedestrian while they encounter each other.
In particular, they are a) GUI-based eHMI with text messages (eHMI-T), b) MUI-based eHMI with neutral voice (eHMI-NV), and c) MUI-based eHMI with affective voice (eHMI-AV) as shown in Table~\ref{Tab:eHMIs}.
The eHMI-T uses a smiley face together with a piece of text to indicate its yielding intention, and uses grateful eyes together with a piece of text to express its appreciation, \ie \textit{thanks}.
Both the eHMI-NV and eHMI-AV have the same GUI design for their eHMI as that for the eHMI-T, but they differ in their VUIs.
To be more specific, the eHMI-NV utilizes a neutral voice~\footnote{The voice is generated by \textit{Microsoft Nanami Online}. The voice pitch was set to -10\% for the neutral voice; and +15\% for the affective voice. The voices for eHMI-NV and eHMI-AV can be found at \url{https://1drv.ms/f/s!AqdIEHyOvvX56E_YDn2VRzSeU1KK}\label{Nanami}} to say ``After you'' and ``Thank you'' to pedestrians, while the eHMI-AV utilizes an affective voice~\textsuperscript{\ref {Nanami}} to express ``Oh please, after you!'' and ``You are so kind. Thank you so much!''.
Both of the voice cues are in Japanese as all participants are native Japanese speakers.

Note that although the basic eHMI can be activated automatically by the APMV's movements (see Table~\ref{Tab:eHMIs_all}), to ensure the performance for communication with pedestrians, a trained experimenter who followed the APMV remotely controled the eHMI and activated the designated GUI and VUIs at a certain moment.

\subsection{Driving Scenarios}

\begin{figure}[t]
  \centering
  \includegraphics[width=0.77\linewidth]{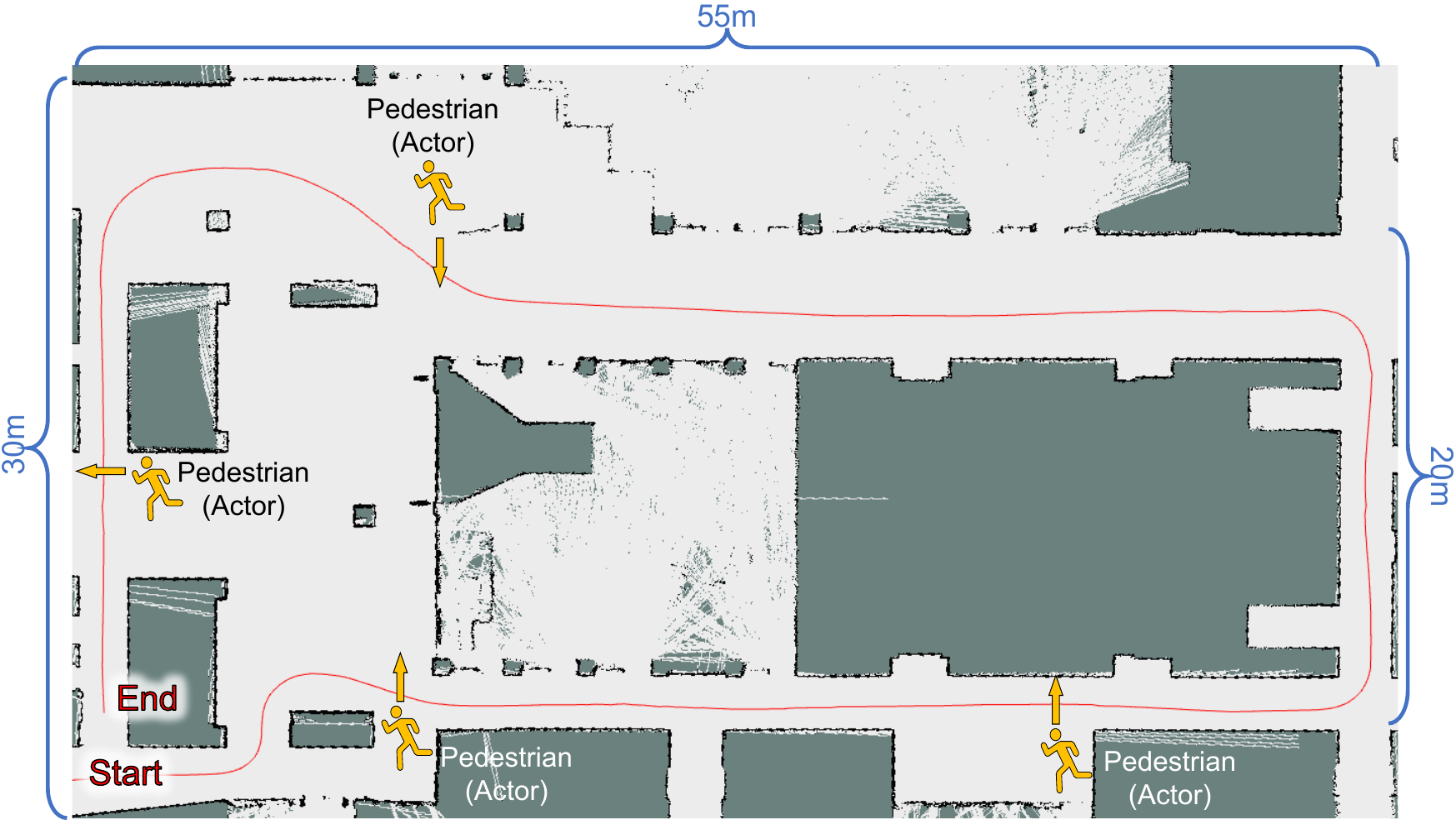}
  \caption{The experimental site is a $55\,m \times 30\,m$ indoor area. In each round, the AVMP's driving route (red line) encounters with pedestrians (actors) four times at the marked positions.
}
  \label{fig:map}
  \end{figure}

For our experimental site, we selected an indoor 55~m $\times$ 30~m area, as shown in Fig.~\ref{fig:map}.
It is a walkway located on the ground floor of the Information Science Complex, at NAIST.
An approximately 150~m long circular driving route has been established at the experimental site. 
This site was chosen based on a common usage scenario in which APMVs drove in a shared indoor area and frequently encountered pedestrians. 
Of these, four specific pedestrians in the interaction scenes were acted by trained actors.
It should be noted that the APMV serves as the subject of communication with pedestrians.
Thus, we asked pedestrians (actors) to face and look at the eyes on the eHMI display, instead of passengers (participants) sitting on the APMV.

\begin{figure}[t]
  \centering
  \includegraphics[width=1\linewidth]{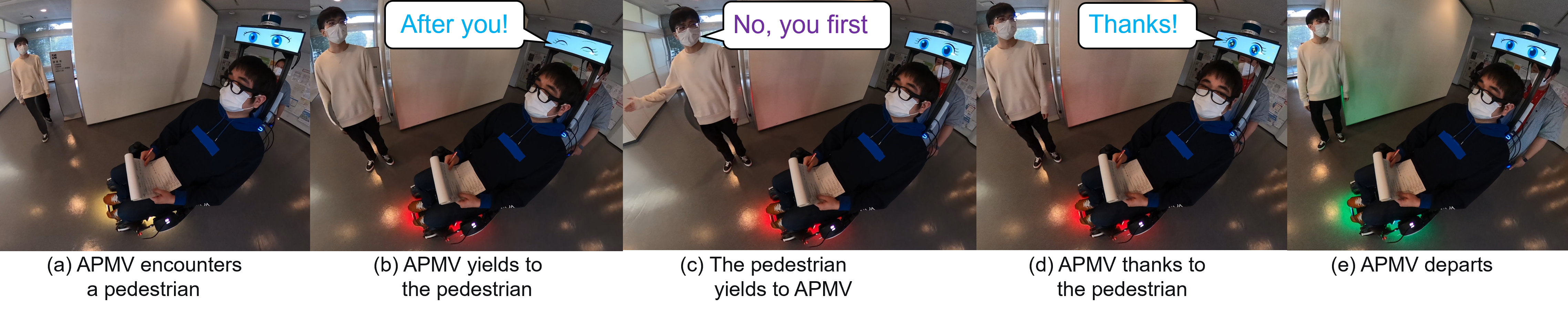}
  \vspace{-7mm}
  \caption{An APMV encounters a pedestrian in an indoor environment and communicates through eHMIs.}
  \label{fig:scene}
  \vspace{-4mm}
\end{figure}

In this site, four similar encounter scenes have been designed along this route (see Fig.~\ref{fig:map}).
The specific interaction scenario at each scene is set up as an APMV and a pedestrian encounter at an intersection with a blind spot/corner.
The APMV stops upon encountering the pedestrian, and likewise, the pedestrian stops synchronously as soon as he/she sees the APMV (see Fig.~\ref{fig:scene}~(a)).
Then, the APMV first indicates its intention to yield to the pedestrian, communicated by the eHMI (see Fig.~\ref{fig:scene}~(b)).
Afterwards, the pedestrian looks at the eye animation showing on the eHMI display and says ``No, after you'' to the APMV (see Fig.~\ref{fig:scene}~(c)).
Finally, the APMV expresses its thanks to the pedestrian via the eHMI  (see Fig.~\ref{fig:scene}~(d)) and then departs (see Fig.~\ref{fig:scene}~(e)).
During the whole encounter scenes, passengers (participants) were asked to ride the APMV in a natural and relaxed way.
Furthermore, there were no restrictions on the passengers while they were communicating spontaneously with pedestrians.

\subsection{Procedure}

First, participants were informed about the experiment content, \ie as a passenger in the APMV to experience the communication between the APMV and pedestrians through different eHMI designs. 
Meanwhile, the three eHMI designs (see Tables~\ref{Tab:eHMIs_all} and \ref{Tab:eHMIs}) were illustrated in detail through a demonstration.
To alleviate any restlessness or nervousness among participants who are unfamiliar with the APMV, we provided an explanation of the principles behind the autonomous driving system and its sensors.
Subsequently, we introduced the questionnaire (see section~\ref{sec:QA}) that participants would need to respond during the experiment.
After signing the informed consent form, participants were instructed to complete a personality inventory before starting the experiment.

\begin{figure}[b]
  \vspace{-4mm}
  \centering
  \includegraphics[width=1\linewidth]{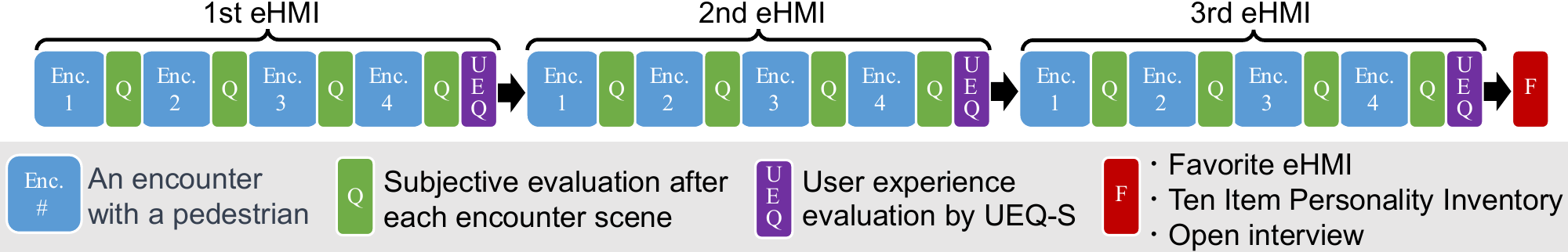}
  \vspace{-5mm}
  \caption{
A schematic diagram of the experimental process.}
  \label{fig:process}
  \vspace{-4mm}
\end{figure}

A schematic diagram of the experimental process is shown in Fig.~\ref{fig:process}.
During the experiment, participants (passengers) sat on the APMV, which autonomously drove around the route (see Fig.~\ref{fig:map}) for three times with a different eHMI in each round, respectively.
It should be noted that the experience order of the three eHMIs and gender of participants were balanced (see Table~\ref{tab:balance}).
Namely, there were six different orders to experience the three eHMIs, and each order was experimented with two participants.
As there are four encounter scenes per round, each participant experienced a total of 12 encounter scenes while riding the APMV as a passenger.
After each encounter scene, participants were required to complete a questionnaire (see section~\ref{sec:QA}) about their subjective feelings of the communication between the APMV (eHMI) and the pedestrians, as perceived from the passenger's perspective.
After each round (including four encounter scenes), participants were asked to complete the user experience questionnaire (see section~\ref{sec:UE}) to assess their experience with the eHMI used.
After all three rounds of experiencing all eHMIs, participants were asked to answer a final question ``Which eHMI is your favorite?'' and give a short open interview.
In the open interview, passengers were free to share their experiential feelings about these three eHMIs and provide the reasons behind their responses.

\begin{figure}[t]
\centering
\captionof{table}{Order of eHMI experience for the 24 participants in the experiment. }
\label{tab:balance}
\renewcommand{\arraystretch}{0.9}
\setlength\tabcolsep{10pt}
\begin{tabular}{@{}c|lclcl@{}}
\toprule
& \multicolumn{5}{c}{Order of experience with eHMIs} \\
 Participants (N=24) & 1st eHMI& & 2nd eHMI& &3rd eHMI\\ \midrule
Male: N=2, Female: N=2 & eHMI-T & $\rightarrow$ & eHMI-NV & $\rightarrow$ & eHMI-AV \\
Male: N=2, Female: N=2 & eHMI-T & $\rightarrow$ & eHMI-AV & $\rightarrow$ & eHMI-NV \\
Male: N=2, Female: N=2 & eHMI-NV & $\rightarrow$ & eHMI-T & $\rightarrow$ & eHMI-AV \\
Male: N=2, Female: N=2 & eHMI-NV & $\rightarrow$ & eHMI-AV & $\rightarrow$ & eHMI-T \\
Male: N=2, Female: N=2 & eHMI-AV & $\rightarrow$ & eHMI-T & $\rightarrow$ & eHMI-NV \\
Male: N=2, Female: N=2 & eHMI-AV & $\rightarrow$ & eHMI-NV & $\rightarrow$ & eHMI-T \\ \bottomrule
\end{tabular}
\vspace{-4mm}
\end{figure}

\section{MEASUREMENT}

\subsection{Subjective evaluation after each encounter scene}
\label{sec:QA}

After each encounter scene, passengers were asked to answer five subjective questions using the 5-point Likert scale, \ie 1=``strongly disagree'', 2=``disagree'', 3=``neutral'', 4=``agree'', and 5=``strongly agree''.
These five questions about how they felt when APMV used eHMI to communicate with pedestrians which are:
\begin{itemize}
\itemsep-0.1em
    \item[Q1:] Was the information conveyed by eHMI insufficient for you?
    \item[Q2:] Was the content of the communication between the eHMI and pedestrians easy to understand?
    \item[Q3:] Did the APMV excessively communicate with pedestrians via the eHMI?
    \item[Q4:] Did you feel awkward during the APMV-pedestrian communication?
    \item[Q5:] Did you worry that the eHMI's performance might attract too much attention from people around you?
\end{itemize}

\subsection{User experience}
\label{sec:UE}

\begin{figure}[h!]
  \centering
  \includegraphics[width=0.7\linewidth]{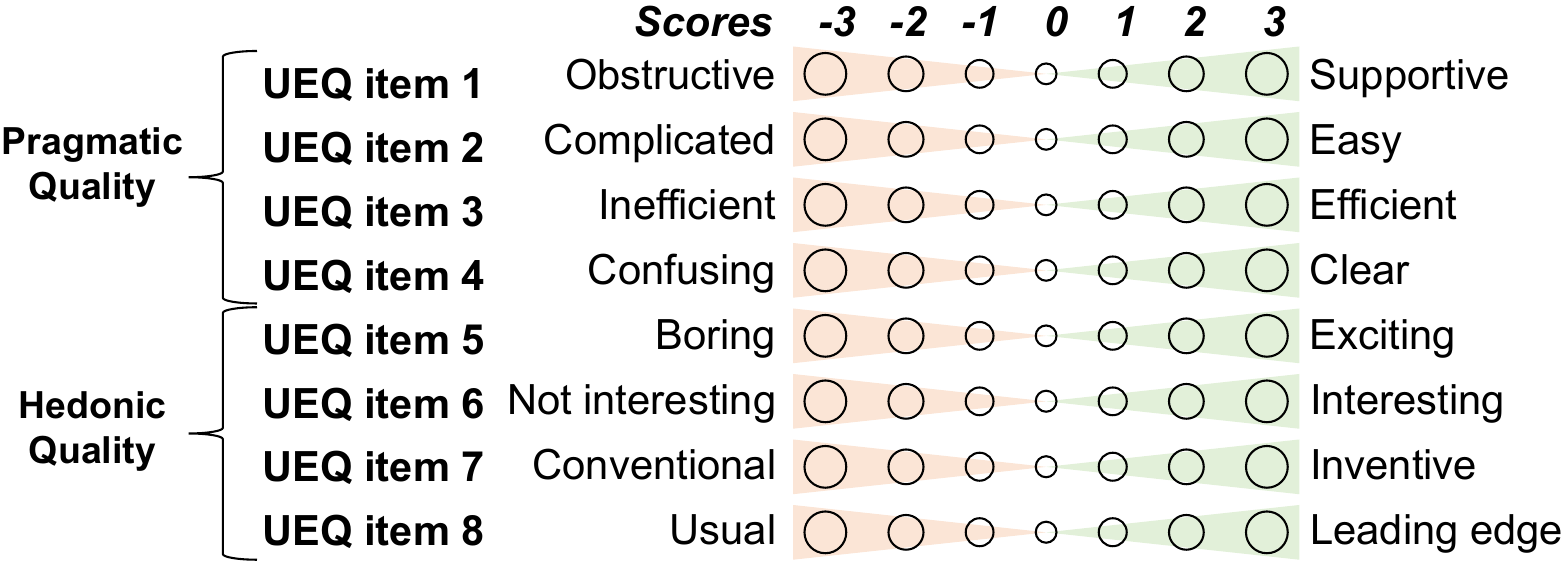}
  \vspace{-2mm}
  \caption{The short version of the user experience questionnaire~(UEQ-S). Note that the Japanese version was used.}
  \label{fig:UEQ-S}
    \vspace{-4mm}
\end{figure}

The short version of the user experience questionnaire~(UEQ-S)~\citep{schrepp2017design} in Japanese~\footnote{The Japanese UEQ-S and the data analysis tool download from \url{https://www.ueq-online.org}\label{ueq-online}} was used to rate the passengers' experience with each design of the eHMIs.
As shown in Fig.~\ref{fig:UEQ-S}, eight items in UEQ-S were analyzed in two user-experience domains, which were counted by:
$\text{Pragmatic Quality}  =\sum_{i=1}^4(\text{UEQ~item}~i)/4$ and $\text{Hedonic Quality}  = \sum_{i=5}^8(\text{UEQ~Item}~i)/4$. 

\subsection{Favorite eHMI}

We analyzed the preference of the three eHMIs in relation to the participants' personality domains. 
Specifically, the participants were divided into three groups based on their responses to the final question ``Which eHMI is your favorite?''
The mean of the five personality domains was compared across the participants in each group.

\subsection{Big 5 Personality Domains}

To find the relationship between the participants' choice of the favorite eHMI and their personality inventory, the Japanese version~\citep{AtsushiOshio2012} of the \textit{Ten Item Personality Inventory} (TIPI)~\citep{gosling2003very} was used.
As shown in Table~\ref{tab:TIPI}, these ten items were rated by the participants using a 7-point scale.

\begin{figure}[bh]
\centering
\footnotesize
\setlength\tabcolsep{3pt}
\renewcommand{\arraystretch}{0.85}
\captionof{table}{Items of \textit{Ten Item Personality Inventory}~\citep{gosling2003very}.}
\label{tab:TIPI}
\begin{tabular}{@{}lccccccc@{}}
\toprule
 & \multicolumn{7}{c}{Scores} \\ \cmidrule(l){2-8} 
 & 1 & 2 & 3 & 4 & 5 & 6 & 7 \\ 
I see myself as: & \begin{tabular}[c]{@{}c@{}}Disagree\\ strongly\end{tabular} & \begin{tabular}[c]{@{}c@{}}Disagree\\ moderately\end{tabular} & \begin{tabular}[c]{@{}c@{}}Disagree\\ a little\end{tabular} & \begin{tabular}[c]{@{}c@{}}Neither agree\\ nor disagree\end{tabular} & \begin{tabular}[c]{@{}c@{}}Agree\\ a little\end{tabular} & \begin{tabular}[c]{@{}c@{}}Agree\\ moderately\end{tabular} & \begin{tabular}[c]{@{}c@{}}Agree\\ strongly\end{tabular} \\ \midrule
TIPI~item~1: Extraverted, Enthusiastic. & $\square$ & $\square$ & $\square$ & $\square$ &$\square$  & $\square$ & $\square$ \\
TIPI~item~2: Critical, Quarrelsome.  & $\square$ & $\square$ & $\square$ & $\square$ &$\square$  & $\square$ & $\square$ \\
TIPI~item~3: Dependable, Self-disciplined.  & $\square$ & $\square$ & $\square$ & $\square$ &$\square$  & $\square$ & $\square$ \\
TIPI~item~4: Anxious, Easily upset.  & $\square$ & $\square$ & $\square$ & $\square$ &$\square$  & $\square$ & $\square$ \\
TIPI~item~5: Open to new experiences, Complex.  & $\square$ & $\square$ & $\square$ & $\square$ &$\square$  & $\square$ & $\square$ \\
TIPI~item~6: Reserved, Quiet.  & $\square$ & $\square$ & $\square$ & $\square$ &$\square$  & $\square$ & $\square$ \\
TIPI~item~7: Sympathetic, Warm.  & $\square$ & $\square$ & $\square$ & $\square$ &$\square$  & $\square$ & $\square$ \\
TIPI~item~8: Disorganized, Careless.  & $\square$ & $\square$ & $\square$ & $\square$ &$\square$  & $\square$ & $\square$ \\
TIPI~item~9: Calm, Emotionally stable.  & $\square$ & $\square$ & $\square$ & $\square$ &$\square$  & $\square$ & $\square$ \\
TIPI~item~10: Conventional, Uncreative.  & $\square$ & $\square$ & $\square$ & $\square$ &$\square$  & $\square$ & $\square$ \\ \bottomrule
\end{tabular}
\end{figure}

As outlined by \cite{AtsushiOshio2012}, the Big 5 personality domains were derived from the scores of TIPI items by aggregating the rating of the corresponding positive item and the residual of the negative item, as following:
\begin{eqnarray} 
\text{Extraversion} & =& \text{TIPI~item~1}+(8-\text{TIPI~item~6}), \nonumber \\ 
\text{Agreeableness} & =& (8-\text{TIPI~item~2})+\text{TIPI~item~7}, \nonumber \\ 
\text{Conscientiousness} &=&\text{TIPI~item~3}+(8-\text{TIPI~item~8}), \nonumber \\ 
\text{Neuroticism} & = &\text{TIPI~item~4}+(8-\text{TIPI~item~9}), \nonumber \\ 
\text{Openness to Experience} & =&\text{TIPI~item~5}+(8-\text{TIPI~item~10}).\nonumber
\end{eqnarray}
Hence, the resulting rating for each personality domain is represented by a discrete value between 1 to 14.

\section{RESULTS}

\subsection{Subjective evaluations after each encounter scene}

\begin{figure}[h!tp]
  \centering
  \includegraphics[width=\linewidth]{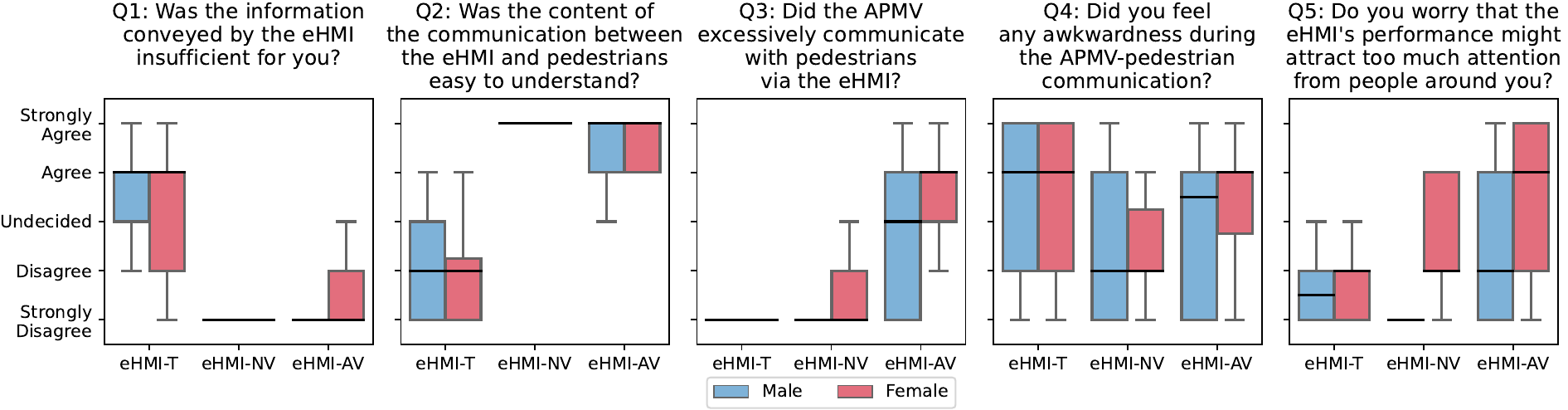}
  \caption{Subjective evaluation results of five questions after each encounter scene, where 1=``strongly disagree'', 2=``disagree'', 3=``neutral'', 4=``agree'', and 5=``strongly agree''.}
  \label{fig:QA}
\vspace{-2mm}
\centering
\small
\setlength\tabcolsep{7.6pt}
\captionof{table}{Two-way mixed ANOVA for each subjective evaluation question. *:$p<.05$, ***:$p<.001$.}
\label{tab:QA_ANOVA}
\begin{tabular}{@{}crrccrrlr}
\toprule
Questions & \multicolumn{1}{c}{Source} & \multicolumn{1}{c}{Sum of Squares} & dof1 & dof2 & \multicolumn{1}{c}{Mean Square} & \multicolumn{1}{c}{\textit{F}-value} & \multicolumn{1}{c}{\textit{p}-value} & \multicolumn{1}{c}{$\eta_p^2$}  \\ \midrule
 & Gender & 0.001 & 1 & 22 & 0.001 & 0.001 & 0.976 & 0.000 \\
Q1 & eHMI & 74.043 & 2 & 44 & 37.022 & 64.816 & 0.000 *** & 0.747 \\
 & Gender~$\times$~eHMI & 0.366 & 2 & 44 & 0.183 & 0.321 & 0.727 & 0.014 \\ \midrule
 & Gender & 0.003 & 1 & 22 & 0.003 & 0.004 & 0.950 & 0.000 \\
Q2 & eHMI & 107.220 & 2 & 44 & 53.610 & 136.115 & 0.000 *** & 0.861 \\
 & Gender~$\times$~eHMI & 0.283 & 2 & 44 & 0.141 & 0.359 & 0.700 & 0.016 \\ \midrule
 & Gender & 2.084 & 1 & 22 & 2.084 & 2.517 & 0.127 & 0.103 \\
Q3 & eHMI & 46.568 & 2 & 44 & 23.284 & 36.449 & 0.000 *** & 0.624 \\
 & Gender~$\times$~eHMI & 2.950 & 2 & 44 & 1.475 & 2.309 & 0.111 & 0.095 \\ \midrule
 & Gender & 1.188 & 1 & 22 & 1.188 & 0.465 & 0.502 & 0.021 \\
Q4 & eHMI & 19.000 & 2 & 44 & 9.500 & 9.974 & 0.000 *** & 0.312 \\
 & Gender~$\times$~eHMI & 0.715 & 2 & 44 & 0.358 & 0.375 & 0.689 & 0.017 \\ \midrule
 & Gender & 11.681 & 1 & 22 & 11.681 & 6.573 & 0.018 * & 0.230 \\
Q5 & eHMI & 17.382 & 2 & 44 & 8.691 & 11.170 & 0.000 *** & 0.337 \\
 & Gender~$\times$~eHMI & 3.465 & 2 & 44 & 1.733 & 2.227 & 0.120 & 0.092 \\ \bottomrule
\end{tabular}
\vspace{2mm}
\centering
\small
\setlength\tabcolsep{7pt}
\captionof{table}{ Post-hoc two-sided pairwise comparisons using the Wilcoxon signed rank test with Benjamini/Hochberg FDR correction for main effect of eHMI for each subjective evaluation question. *:$p<.05$, **:$p<.01$, ***:$p<.001$.}
\label{tab:QA_post-hoc}
\begin{tabular}{@{}cllccccrlr@{}}
\toprule
Qustions & \multicolumn{1}{c}{A} & \multicolumn{1}{c}{B} & mean(A) & std(A) & mean(B) & std(B) & \multicolumn{1}{c}{\textit{W}-value} & \multicolumn{1}{c}{\textit{p}-adj} & \multicolumn{1}{r}{hedges' \textit{g}} \\ \midrule
 & eHM-T & eHM-NV & 3.479 & 1.058 & 1.344 & 0.725 & 0.000 & 0.000 *** & 2.316 \\
Q1 & eHM-T & eHM-AV & 3.479 & 1.058 & 1.312 & 0.586 & 0.000 & 0.000 *** & 2.492 \\
 & eHM-NV & eHM-AV & 1.344 & 0.725 & 1.312 & 0.586 & 24.500 & 0.798 & 0.047 \\ \midrule
 & eHM-T & eHM-NV & 2.115 & 0.997 & 4.719 & 0.614 & 0.000 & 0.000 *** & -3.094 \\
Q2 & eHM-T & eHM-AV & 2.115 & 0.997 & 4.688 & 0.456 & 0.000 & 0.000 *** & -3.264 \\
 & eHM-NV & eHM-AV & 4.719 & 0.614 & 4.688 & 0.456 & 20.500 & 0.503 & 0.057 \\ \midrule
 & eHM-T & eHM-NV & 1.281 & 0.558 & 1.396 & 0.546 & 35.500 & 0.505 & -0.204 \\
Q3 & eHM-T & eHM-AV & 1.281 & 0.558 & 3.042 & 1.274 & 16.000 & 0.000 *** & -1.761 \\
 & eHM-NV & eHM-AV & 1.396 & 0.546 & 3.042 & 1.274 & 0.000 & 0.000 *** & -1.652 \\ \midrule
 & eHM-T & eHM-NV & 3.656 & 1.318 & 2.406 & 1.075 & 20.000 & 0.003 ** & 1.022 \\
Q4 & eHM-T & eHM-AV & 3.656 & 1.318 & 3.156 & 1.206 & 80.000 & 0.134 & 0.389 \\
 & eHM-NV & eHM-AV & 2.406 & 1.075 & 3.156 & 1.206 & 31.000 & 0.009 ** & -0.646 \\ \midrule
 & eHM-T & eHM-NV & 1.896 & 0.918 & 1.938 & 1.046 & 60.000 & 0.696 & -0.042 \\
Q5 & eHM-T & eHM-AV & 1.896 & 0.918 & 2.958 & 1.382 & 27.000 & 0.006 ** & -0.891 \\
 & eHM-NV & eHM-AV & 1.938 & 1.046 & 2.958 & 1.382 & 31.500 & 0.006 ** & -0.819 \\ \bottomrule
\end{tabular}
\end{figure}

A summary of the answers to the five subjective evaluations after each encounter scene, \ie Q1 to Q5 described in section~\ref{sec:QA}, reported by the 24 participants is shown in Fig.~\ref{fig:QA}. 
As shown in Table~\ref{tab:QA_ANOVA}, a two-way mixed ANOVA revealed that there was not a statistically significant interaction between the effects of gender and eHMI on Q1-Q5.
Additionally, it showed a main effect of gender was significant on Q5 ($p<.05$) only, indicating that females were significantly more concerned about eHMI attracting too much attention from people around them than males.
Moreover, the two-way mixed ANOVA also reported that main effects of eHMI were significant on Q1 to Q5 ($p<.001$), respectively.

Post-hoc pairwise comparisons of eHMIs for Q1 to Q5 using the Wilcoxon signed-rank test with BH-FDR correction are presented in Table~\ref{tab:QA_post-hoc}.
For Q1, passengers perceived that eHMI-T conveyed information significantly insufficient compared to eHMI-NV ($p<.001$) and eHMI-AV ($p<.001$).
Similarly, from the results of Q2, passengers agreed that the communication contents were significantly easier to understand when using eHMI-NV ($p<.001$) and eHMI-AV ($p<.001$) compared to using eHMI-T.
However, in Q3, passengers also significantly agreed that eHMI-AV exhibited excessive communication compared to eHMI-T ($p<.001$) and eHMI-NV ($p<.001$).
Comparison results from Q4 indicated that passengers felt significantly more discomfort when experiencing both eHMI-T ($p<.01$) and eHMI-AV ($p<.01$) compared to using eHMI-NV.
Regarding Q5, passengers perceived that when experiencing eHMI-AV, they significantly worried about attracting too much attention from people compared to eHMI-T ($p<.01$) and eHMI-NV ($p<.01$).

\subsection{User experience}
\label{sec:UEQ_results}

The results of the 24 participants answering the UEQ-S for three eHMIs were calculated into two user-experience domains, \ie pragmatic quality and hedonic quality.
Cronbach's alpha in pragmatic quality items, \ie UEQ-Item~1 to 4, is 0.78, and
Cronbach's alpha in hedonic quality items, \ie UEQ-Item~5 to 8, is 0.92.

\begin{figure}[h!t]
  \centering
  \includegraphics[width=0.76\linewidth]{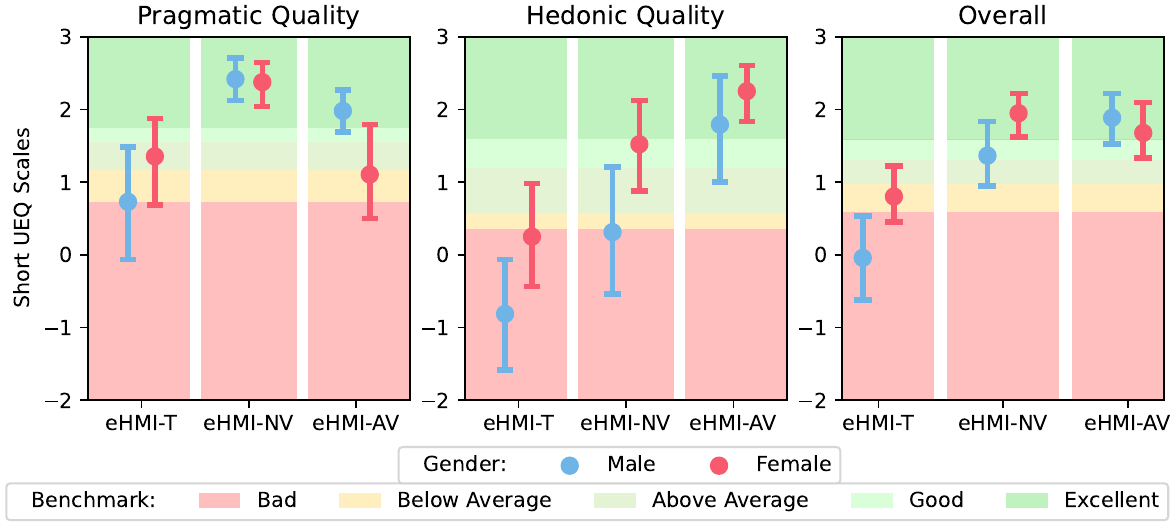}
\captionof{figure}{Results of UEQ-S for three types of eHMIs with the UEQ benchmark. Error bars show their confidence intervals.}
  \label{fig:UEQ-S_result}
\end{figure}

The evaluation results of user experience for pragmatic quality and hedonic quality of three eHMIs by male and female participants are shown in Fig.~\ref{fig:UEQ-S_result}.
The average values of pragmatic quality and hedonic quality were calculated as the overall user experience results, which are shown in Fig.~\ref{fig:UEQ-S_result}.
Background colors of each domain in Fig.~\ref{fig:UEQ-S_result} represent the user experience benchmarks obtained from a dataset including over 400 studies that used the UEQ to evaluate different products~\citep{hinderks2018benchmark}.

The two-way mixed ANOVA results in Table~\ref{tab:UE_ANOVA} indicated significant effects of eHMI on various aspects of user experience.
Specifically, eHMI significantly influenced the pragmatic quality ($p<.001$), hedonic quality ($p<.001$), and the overall user experience ($p<.001$). 
In terms of gender, the main effect was significant only on the hedonic quality ($p<.05$), indicating that the influence of eHMI on the hedonic quality of user experience may also depending on the gender.
The two-way mixed ANOVA further revealed a significant interaction between the effects of gender and eHMI on the pragmatic quality ($p<.05$) and the overall user experience ($p<.05$).
However, this interaction was not significant for the hedonic quality.

Table~\ref{tab:UE_simple_effect_1} presents the simple main effects of gender within eHMIs on the pragmatic quality and overall user experience by using two-sided t-tests with Holm correction.
For all eHMIs, no significant differences were found in the pragmatic quality and overall user experience due to gender differences.

Table~\ref{tab:UE_simple_effect_2} presents the the simple main effects of eHMIs within gender on the pragmatic quality and overall user experience by using by two-sided paired t-tests with Holm correction.
For the pragmatic quality, both females and males rated eHMI-NV significantly higher than eHMI-T (Female: $p<.05$, Male: $p<.01$), as well as eHMI-NV significantly higher than eHMI-AV (Female: $p<.05$, Male: $p<.05$).
However, a gender difference was observed in the pragmatic quality of eHMI-AV and eHMI-T.
Only males rated eHMI-AV significantly higher than eHMI-T ($p<.05$) in terms of pragmatic quality, while females did not perceive a significant difference between the two.
In terms of overall user experience, both males and females rated eHMI-T significantly lower than eHMI-NV (Female: $p<.001$, Male: $p<.001$) and eHMI-AV (Female: $p<.05$, Male: $p<.001$). 
However, the significant difference in overall user experience between eHMI-NV and eHMI-AV was only observed in males ($p<.05$), not in females.

The post-hoc pairwise comparisons for main effect of eHMIs, conducted using two-sided t-tests with Holm correction across different user experience domains, are detailed in Table~\ref{tab:UE_post-hoc}.
In which, eHMI-NV significantly outperforms both eHMI-T ($p<.001$) and eHMI-AV ($p<.001$) in the pragmatic quality, but there is no significant difference between eHMI-T and eHMI-AV.
For hedonic quality, eHMI-T significantly underperforms both eHMI-NV ($p<.001$) and eHMI-AV ($p<.001$), while eHMI-AV significantly outperforms eHMI-NV ($p<.001$).
When considering the overall user experience, taking into account both pragmatic quality and hedonic quality, eHMI-T significantly underperforms both eHMI-NV ($p<.001$) and eHMI-AV ($p<.001$). However, there is no significant difference between eHMI-NV and eHMI-AV.

\begin{figure}[h!p]
\centering
\small
\renewcommand{\arraystretch}{0.9}
\setlength\tabcolsep{6pt}
\captionof{table}{Two-way mixed-ANOVA for user experience (UE) domains. *:$p<.05$,**:$p<.01$, ***:$p<.001$.}
\label{tab:UE_ANOVA}
\begin{tabular}{@{}crrccrrlr}
\toprule
\multicolumn{1}{c}{UE domains} & \multicolumn{1}{c}{Source} & \multicolumn{1}{c}{Sum of Aqueres} & dof1 & dof2 & \multicolumn{1}{c}{Mean Square } & \multicolumn{1}{c}{\textit{F}-value} & \multicolumn{1}{c}{\textit{p}-value} & \multicolumn{1}{c}{$\eta_p^2$}  \\ \midrule
 & Gender & 0.170 & 1 & 22 & 0.170 & 0.128 & 0.724 & 0.006\\
Pragmatic Quality & eHMI & 22.507 & 2 & 44 & 11.253 & 15.820 & 0.000 *** & 0.418\\
 & Gender~$\times$~eHMI & 6.778 & 2 & 44 & 3.389 & 4.764 & 0.013 * & 0.178\\ \midrule
& Gender & 14.897 & 1 & 22 & 14.897 & 4.509 & 0.045 * & 0.170\\
Hedonic Quality& eHMI & 63.630 & 2 & 44 & 31.815 & 33.822 & 0.000 *** & 0.606\\
 & Gender~$\times$~eHMI & 1.898 & 2 & 44 & 0.949 & 1.009 & 0.373 & 0.044\\ \midrule
 & Gender & 2.971 & 1 & 22 & 2.971 & 2.814 & 0.108 & 0.113\\
Overall & eHMI & 28.855 & 2 & 44 & 14.427 & 38.593 & 0.000 *** & 0.637\\
 & Gender~$\times$~eHMI & 3.603 & 2 & 44 & 1.801 & 4.819 & 0.013 * & 0.180\\ \bottomrule
\end{tabular}

\centering
\captionof{table}{Simple main effects of gender within eHMIs by t-tests (two-sided) with Holm correction for user experience (UE) domains.}
\label{tab:UE_simple_effect_1}
\small
\renewcommand{\arraystretch}{0.95}
\setlength\tabcolsep{4.2pt}
\begin{tabular}{@{}clllrrrrrclr@{}}
\toprule
UE domains & \multicolumn{1}{c}{eHMIs} & \multicolumn{1}{c}{A} & \multicolumn{1}{c}{B} & \multicolumn{1}{c}{mean(A)} & \multicolumn{1}{c}{std(A)} & \multicolumn{1}{c}{mean(B)} & \multicolumn{1}{c}{std(B)} & \multicolumn{1}{c}{\textit{T}-value} & dof & \multicolumn{1}{c}{\textit{p}-adj} & \multicolumn{1}{c}{cohen's \textit{d}} \\ \midrule
 & eHMI-T & Female & Male & 1.354 & 1.014 & 0.729 & 1.463 & 1.216 & 22 & 0.474 & 0.496 \\ \cmidrule(l){2-12} 
 Pragmatic Quality& eHMI-NV & Female & Male & 2.375 & 0.549 & 2.417 & 0.557 & -0.185 & 22 & 0.855 & -0.075 \\ \cmidrule(l){2-12} 
 & eHMI-AV & Female & Male & 1.104 & 1.180 & 1.979 & 0.579 & -2.307 & 22 & 0.093 & -0.942 \\ \midrule
 & eHMI-T & Female & Male & 0.802 & 0.712 & -0.042 & 1.084 & 1.216 & 22 & 0.103 & 0.920 \\ \cmidrule(l){2-12} 
Overall & eHMI-NV & Female & Male & 1.948 & 0.540 & 1.365 & 0.842 & -0.185 & 22 & 0.111 & 0.825 \\ \cmidrule(l){2-12} 
 & eHMI-AV & Female & Male & 1.677 & 0.698 & 1.885 & 0.662 & -2.307 & 22 & 0.461 & -0.306 \\ \bottomrule
\end{tabular}

\centering
\captionof{table}{Simple main effects of eHMIs within gender by paired t-tests (two-sided) with Holm correction for user experience (UE) domains.  *:$p<.05$, **:$p<.01$, ***:$p<.01$.}
\label{tab:UE_simple_effect_2}
\small
\renewcommand{\arraystretch}{0.95}
\setlength\tabcolsep{2.5pt}
\begin{tabular}{@{}clllrrrrrclr@{}}
\toprule
UE domains & \multicolumn{1}{c}{Genders} & \multicolumn{1}{c}{A} & \multicolumn{1}{c}{B} & \multicolumn{1}{c}{mean(A)} & \multicolumn{1}{c}{std(A)} & \multicolumn{1}{c}{mean(B)} & \multicolumn{1}{c}{std(B)} & \multicolumn{1}{c}{\textit{T}-value} & dof & \multicolumn{1}{c}{\textit{p}-adj} & \multicolumn{1}{c}{cohen's \textit{d}} \\ \midrule
&  & eHMI-AV & eHMI-NV & 1.104 & 1.180 & 2.375 & 0.549 & -3.302 & 11 & 0.022 * & -1.381 \\
& Female & eHMI-AV & eHMI-T & 1.104 & 1.180 & 1.354 & 1.014 & -0.504 & 11 & 0.624 & -0.227 \\
\multirow{2}{*}{Pragmatic Quality}&  & eHMI-NV & eHMI-T & 2.375 & 0.549 & 1.354 & 1.014 & 3.431 & 11 & 0.022 * & 1.252 \\ \cmidrule(l){2-12} 
 &  & eHMI-AV & eHMI-NV & 1.979 & 0.579 & 2.417 & 0.557 & -3.339 & 11 & 0.022 * & -0.770 \\
 & Male & eHMI-AV & eHMI-T & 1.979 & 0.579 & 0.729 & 1.463 & 3.927 & 11 & 0.012 * & 1.123 \\
 &  & eHMI-NV & eHMI-T & 2.417 & 0.557 & 0.729 & 1.463 & 5.089 & 11 & 0.002 ** & 1.524 \\ \midrule
 &  & eHMI-AV & eHMI-NV & 1.677 & 0.698 & 1.948 & 0.540 & -1.309 & 11 & 0.217 & -0.434 \\
 & Female & eHMI-AV & eHMI-T & 1.677 & 0.698 & 0.802 & 0.712 & 2.843 & 11 & 0.048 * & 1.241 \\
\multirow{2}{*}{Overall} &  & eHMI-NV & eHMI-T & 1.948 & 0.540 & 0.802 & 0.712 & 5.714 & 11 & 0.001 *** & 1.814 \\ \cmidrule(l){2-12} 
 &  & eHMI-AV & eHMI-NV & 1.885 & 0.662 & 1.365 & 0.842 & 2.832 & 11 & 0.048 * & 0.688 \\
 & Male & eHMI-AV & eHMI-T & 1.885 & 0.662 & -0.042 & 1.084 & 5.856 & 11 & 0.001 *** & 2.146 \\
 &  & eHMI-NV & eHMI-T & 1.365 & 0.842 & -0.042 & 1.084 & 6.051 & 11 & 0.000 *** & 1.449 \\ \bottomrule
\end{tabular}

\centering
\captionof{table}{Post-hoc pairwise comparisons by paired t-tests (two-sided) with Holm correction for user experience (UE) domains.  *:$p<.05$, **:$p<.01$, ***:$p<.01$.}
\label{tab:UE_post-hoc}
\small
\renewcommand{\arraystretch}{0.95}
\setlength\tabcolsep{4.4pt}
\begin{tabular}{@{}cllrrrrrclr@{}}
\toprule
UE domains & \multicolumn{1}{c}{A} & \multicolumn{1}{c}{B} & mean(A) & std(A) & mean(B) & std(B) & \textit{T}-value & dof & \multicolumn{1}{c}{\textit{p}-adj} & cohen's \textit{d} \\ \midrule
 & eHMI-AV & eHMI-NV & 1.542 & 1.013 & 2.396 & 0.541 & -3.937 & 23 & 0.001 *** & -1.052 \\
Pragmatic Quality & eHMI-AV & eHMI-T & 1.542 & 1.013 & 1.042 & 1.272 & 1.525 & 23 & 0.141 & 0.435 \\
 & eHMI-NV & eHMI-T & 2.396 & 0.541 & 1.042 & 1.272 & 5.922 & 23 & 0.000 *** & 1.385 \\ \midrule
 & eHMI-AV & eHMI-NV & 2.021 & 1.096 & 0.917 & 1.542 & 4.096 & 23 & 0.000 *** & 0.825 \\
Hedonic Quality & eHMI-AV & eHMI-T & 2.021 & 1.096 & -0.281 & 1.453 & 6.683 & 23 & 0.000 *** & 1.789 \\
 & eHMI-NV & eHMI-T & 0.917 & 1.542 & -0.281 & 1.453 & 5.716 & 23 & 0.000 *** & 0.800 \\ \midrule
 & eHMI-AV & eHMI-NV & 1.781 & 0.674 & 1.656 & 0.753 & 0.788 & 23 & 0.438 & 0.175 \\
Overall & eHMI-AV & eHMI-T & 1.781 & 0.674 & 0.380 & 0.995 & 5.693 & 23 & 0.000 *** & 1.649 \\
 & eHMI-NV & eHMI-T & 1.656 & 0.753 & 0.380 & 0.995 & 8.366 & 23 & 0.000 *** & 1.446 \\ \bottomrule
\end{tabular}
\end{figure}

\newpage
\subsection{Favorite eHMI}
\label{sec:favorite_eHMI}

\begin{figure}[h!b]
\centering
\includegraphics[width=0.9\linewidth]{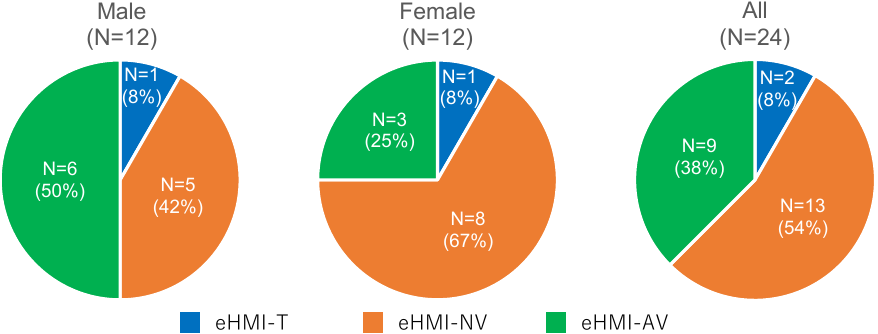}
\caption{Numbers of male, female as well as all participants who selected each type of favorite eHMI.}
\label{fig:favorite_eHMI}
\end{figure}

After experiencing all eHMIs, participants were asked to select the favorite eHMI.
Figure~\ref{fig:favorite_eHMI} shows the number of female, male as well as all participants who selected each type of favorite eHMI.
Fisher's exact test (two-sided) was used to determine if there was a significant association between two genders and three favorite eHMIs.
Its result indicated that there was not a statistically significant association between the favorite eHMIs and genders ($p =0.680$).

\subsection{Big 5 personality domains}
\label{sec:Big5}

\begin{figure}[h!tb]
\centering
\includegraphics[width=1\linewidth]{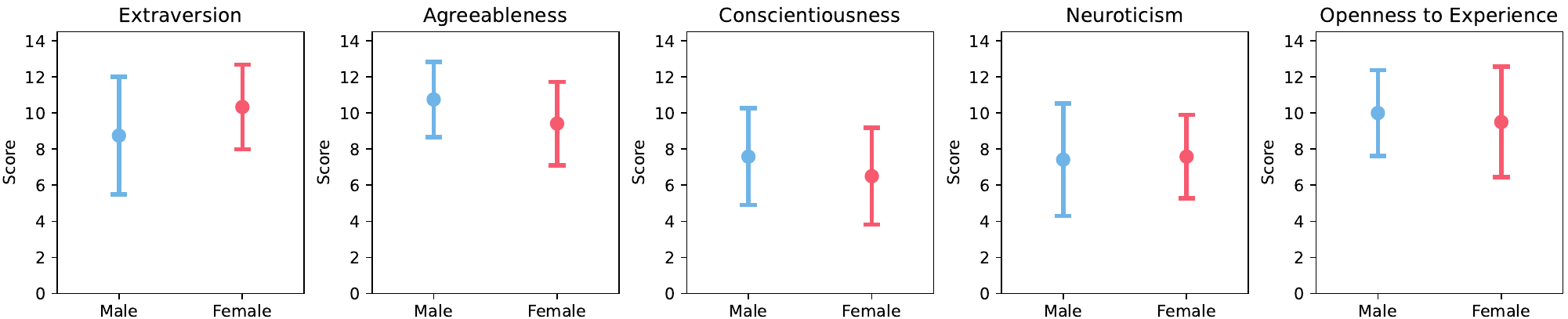}
\captionof{figure}{Distribution of the big 5 personality domains for male and female participants. Dots indicate mean scores and error bars indicate standard deviations.}
  \label{fig:Big5}

\vspace{3mm}

\centering
\captionof{table}{The Big 5 personality domains of 12 males and 12 females participants. A two-sided t-test was used to examine each personality domain between genders.}
 \label{tab:Big5}
\centering
\begin{tabular}{@{}lrrrrrrrrrr@{}}
\toprule
 & \multicolumn{2}{c}{Female} & \multicolumn{2}{c}{Male} & & \multicolumn{5}{c}{t-test (Female,Male)} \\ \cmidrule(l){2-3} \cmidrule(l){4-5} \cmidrule(l){7-11} 
\multicolumn{1}{c}{Personality domains} & \multicolumn{1}{c}{mean} & \multicolumn{1}{c}{std.} & \multicolumn{1}{c}{mean} & \multicolumn{1}{c}{std.} & & \multicolumn{1}{c}{\textit{T}-value} & \multicolumn{1}{c}{dof} & \multicolumn{1}{c}{\textit{p-value}} & \multicolumn{1}{c}{BF10} & \multicolumn{1}{c}{hedges' \textit{g}} \\ \midrule
Extraversion & 10.333 & 2.348 & 8.750 & 3.251 &  &1.368 & 22 & 0.185 & 0.729 & 0.539 \\
Agreeableness & 9.417 & 2.314 & 10.750 & 2.094 &  &-1.480 & 22 & 0.153 & 0.814 & -0.583 \\
Conscientiousness & 6.500 & 2.680 & 7.583 & 2.678 &  &-0.990 & 22 & 0.333 & 0.533 & -0.390 \\
Neuroticism & 7.583 & 2.314 & 7.417 & 3.118 &  &0.149 & 22 & 0.883 & 0.376 & 0.059 \\
Openness to Experience & 9.500 & 3.060 & 10.000 & 2.374 &  &-0.447 & 22 & 0.659 & 0.402 & -0.176 \\ \bottomrule
\end{tabular}
\end{figure}

The results of the Big 5 personality domains for 12 males and 12 females participants are shown in Fig.~\ref{fig:Big5}.
For each personality domain, a t-test (two-sided) was used to test for differences between male and female.
Table~\ref{tab:Big5} shows that there were no significant differences between males and females in all five personality domains.

\subsection{Relation between the favorite eHMI and personality}
\label{sec:big5-eHMI}

In sections~\ref{sec:favorite_eHMI} and~\ref{sec:Big5}, no significant differences were observed between males and females in their choice of favorite eHMI and their Big 5 personality domains.
Therefore, in this section, the analysis of the relationship between participants' choice of favorite eHMI and their personality domains did not take into account gender differences.

\begin{figure}[h!tb]
\centering
\includegraphics[width=1\linewidth]{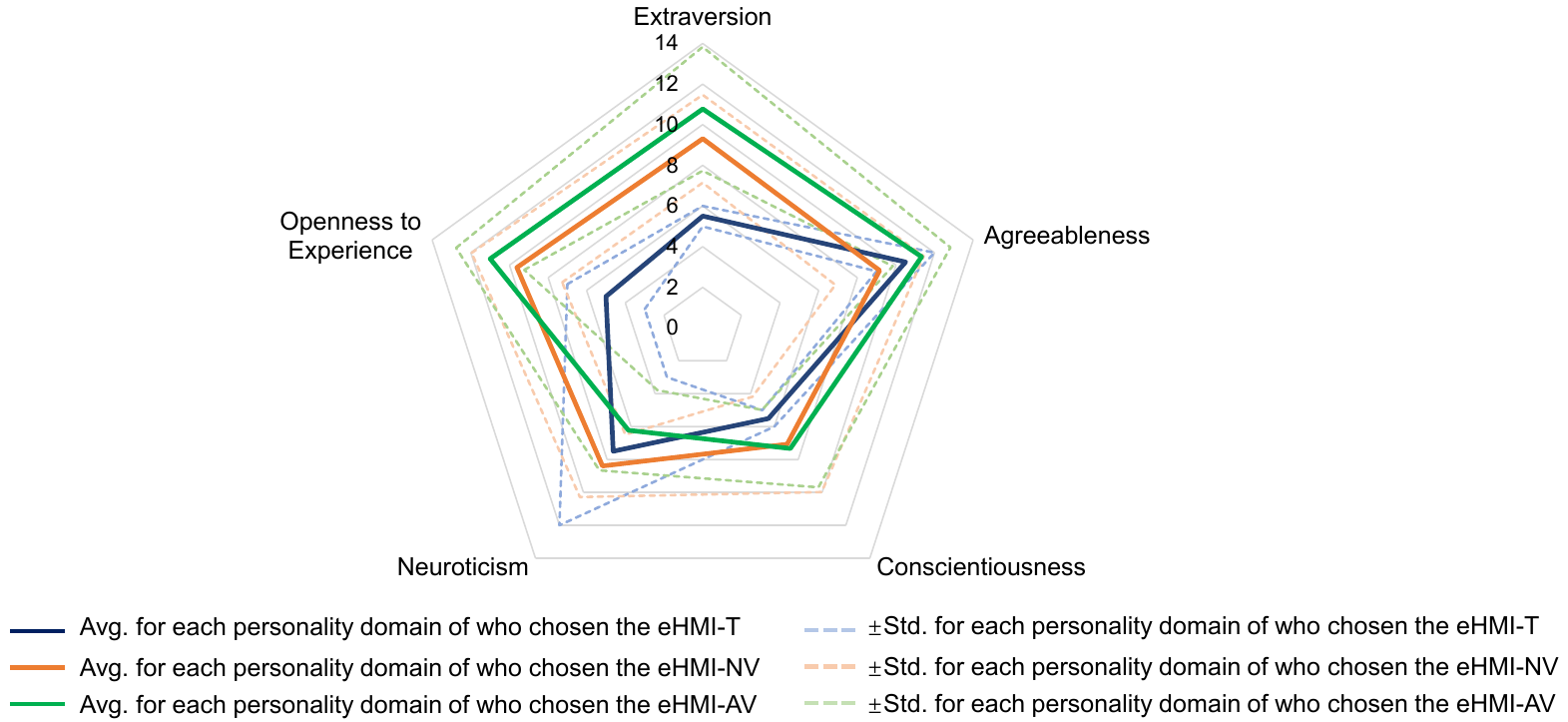}
\captionof{figure}{ Big 5 personality domains of the participants based on their favorite eHMI. The solid line represents the mean, while the dashed line represents the standard deviation.}
  \label{fig:FQ1-TITP}

\end{figure}

Figure~\ref{fig:FQ1-TITP} shows the differences of the Big 5 personality domains of the participants based on their favorite eHMI.
In which, participants with lower scores in extraversion, openness to experience and conscientiousness tended to favor eHMI-T, while those with higher scores in those domains tended to prefer eHMI-AV.

It is worth noting that although the average neuroticism of passengers who chose eHMI-T is between the average neuroticism of people who chose eHMI-AV and eHMI-NV, its standard deviation was large.
This is primarily due to the small sample size of passengers who chose eHMI-T, specifically, one male and one female.

Compared to those who favored eHMI-AV, individuals who preferred eHMI-NV exhibited lower levels of extraversion, openness to experience, and agreeableness, while demonstrating a slightly higher degree of neuroticism.
This suggests that different eHMIs may appeal to different personality types.

\section{DISCUSSION}
\subsection{User Experience of APMV Passengers on the silent GUI-based eHMI}

The subjective evaluation results of Q1 and Q2 in Fig.~\ref{fig:QA} and Table~\ref{tab:QA_post-hoc} show that, compared with the MUI-based eHMIs (\ie eHMI-NV and eHMI-AV), the passengers got insufficient information from eHMI-T, which hindered their situation awareness and made them more difficult to understand the communication between APMV and pedestrians.
In post-experiment interviews, participants made the same comments as ``\textit{without the sound} (\ie eHMI-T)\textit{, I don't know what is being displayed. The pedestrians are talking by themselves and I don't know what kind of communication they are having.}''

Moreover, this lack of understand led passengers to feelings of awkwardness, as indicated in Q4. 
Also, from the interview, almost all passengers reported that the silence time when eHMI-T conveyed information to pedestrians was awkward.
For example, ``\textit{there was a long period of silence in the encounter, which caused a strong sense of awkwardness.}'' Moreover, some passengers also reported feeling ``\textit{left out when eHMI-T and pedestrians communicated.}''

Compared to other MUI-based eHMIs, the above issues resulted in significantly lower ratings of the eHMI-T’s pragmatic quality and hedonic quality of user experience by both male and female passengers (see Fig.~\ref{fig:UEQ-S_result}).
Particularly when compared to the benchmark in~\citep{hinderks2018benchmark}, the hedonic quality of the eHMI-T was already categorized in the bad range.
This suggests that utilizing the voice cues may offer a more engaging and enjoyable experience for passengers compared to relying solely on a silent GUI-based eHMI.

According to the above research results, the answer to \textbf{RQ~1} is that from the perspective of passengers, the silent GUI-based eHMI may not be suitable for APMV because of insufficient information, difficulties in understanding and the feeling of awkwardness during the APMV-pedestrian communications.

\subsection{User Experience of APMV Passengers on the MUI-based eHMIs}

The findings from Q1 and Q2, illustrated in Fig.~\ref{fig:QA} and Table~\ref{tab:QA_post-hoc}, reveal that communication with pedestrians through two MUI-based eHMIs, specifically eHMI-NV and eHMI-AV, enabled the sharing of information with passengers via voice cues, thereby facilitating passengers' understanding of the communication.

However, as shown in Fig.~\ref{fig:QA} and Table~\ref{tab:QA_post-hoc}, the results of Q3 and Q4 indicated that passengers felt the eHMI-AV communicated excessively compared to the eHMI-NV, leading to significant feeling of awkwardness for passengers experiencing the communication between eHMI-AV and pedestrians.
From the interviews, we found that there are differences in the experience of male and female passengers with the eHMI-AV.
Specifically, a few male passengers felt that the eHMI-AV gave too much information compared to the eHMI-NV, \eg ``\textit{I preferred eHMI-NV over eHMI-AV because it provided the minimum necessary communication.}''
However, half of the female passengers said that ``the eHMI-AV was too casual, especially when the encountering pedestrian who was a stranger.''

The results from Q5, as shown in Table~\ref{tab:QA_post-hoc}, indicated that passengers felt significantly more self-conscious about attracting attention from those around them when using the eHMI-AV compared to the eHMI-NV.
Interestingly, Fig.~\ref{fig:QA} revealed a gender disparity in the responses to Q5 for eHMI-NV.
Specifically, male passengers were not concerned about eHMI-NV drawing too much attention from people around them, whereas female passengers expressed more concern about this issue. 
This suggests that gender may play a role in shaping passengers' perceptions and experiences when using these interfaces.
This finding can be explained by the conclusion in \citet{turk1998investigation}, which reported that in some public places, females were significantly greater fear than males being the center of attention.
Similarly, \citep{zentner2023cultural} also found that females have higher levels of social anxiety than males.

The above conclusions can also be reflected in the UEQ results described in the section~\ref{sec:UEQ_results}.
The results presented in Fig.\ref{fig:UEQ-S_result} and Table~\ref{tab:UE_post-hoc} suggest that eHMI-NV was found to have significantly better pragmatic quality than eHMI-AV.
This is consistent with the results of Q3, Q4, Q5 in Fig.~\ref{fig:QA} and Table~\ref{tab:QA_post-hoc}, \ie the eHMI-NV accomplishes negotiating with pedestrians in the most concise communication possible, without the perceived excessive communication that causes passengers to feel awkward, as is the case with the eHMI-AV.
In the interview we got similar results, such as ``\textit{I felt that they communicated better with voice than with text} (eHMI-T), \textit{but I preferred eHMI-NV because it communicated the minimum required compared to eHMI-AV.}''

However, from the perspective of hedonic quality, eHMI-AV is considered significantly better than eHMI-NV, as shown in Fig.\ref{fig:UEQ-S_result} and Table~\ref{tab:UE_post-hoc}.
In the interview, some passengers told that ``\textit{I like the eHMI-AV better than the eHMI-NV because I feel that the eHMI-AV is more communicative,}'' as well as ``\textit{I think the robot character} (\ie APMV's eHMI) \textit{should have its own personality.}''
Above results are in line with \cite{lee2019autonomous}, where the more anthropomorphic voice was favored and rated higher in trust, pleasure, and sense of control than a voice-command interface. 
Besides, the speech embodiment of more conversational speech was evaluated as more warm and social presence \cite{wang2021vehicle}.

Therefore, the answer to\textbf{ RQ~2} is that MUI-based eHMIs with voice cues, \ie the eHMI-NV and eHMI-AV, both significantly enhance the passenger's user experience in APMV and pedestrian interactions.
Moreover, it should be noted that each of these two MUI-based eHMIs has its own advantages, i.e., eHMI-NV has an advantage in pragmatic quality, while eHMI-AV has an advantage in hedonic quality.
This differences in pragmatic and hedonic qualities between eHMI-NV and eHMI-AV aligns with the findings from a study on human-robot interaction by \citet{ullrich2017robot}.
They reported that the neutral robot personality was rated best in a goal-oriented situation, while the positive robot personality was preferred in experience-oriented scenarios.
Above results highlight the importance of considering both pragmatic and hedonic quality in the design of eHMIs, and the potential benefits of using voice cues to enhance passengers' user experiences.

\subsection{eHMI designs for personality}

To answer the RQ~3, the relationship between participants' personalities and their favorite eHMI was discussed.
As results disrupted in section~\ref{sec:big5-eHMI}, this study found that individuals with different personality traits tend to prefer different types of eHMI.
Specifically, those with lower scores in extraversion, openness to experience, and conscientiousness were more inclined to favor eHMI-T, while those with higher scores in these domains tended to prefer eHMI-AV.
This result suggests a potential correlation between personality traits and the choice of eHMI, which could have implications for designing interfaces that cater to the preferences and needs of different personality types.
This result aligns with the conclusion drawn in~\citep{tapus2008user, sarsam2018towards}.
In interviews, 20 out of 24 passengers thought it was important that the eHMI was designed to fit their personality.
A number of representative sentiments were reported, such as
``\textit{If the APMV can reflect the passenger’s personality, that’s fine. But if it deviates from the passenger’s personality, I will feel strange}'', 
``\textit{I think it’s better to match my personality because I feel like it is a part of me}'' as well as ``\textit{I think this is very important because if our personalities don't match well, I might feel it's overly polite, or like my feelings aren't being properly responded to.}''
Moreover, two passengers thought that matching personalities is important, but they have concerns, such as 
``\textit{As a user, I think it’s better if it matches but I don’t like attracting too much attention.}''
The remaining two passengers made it clear that they did not think the eHMI needed to be designed to fit their personalities, for example
``\textit{I don’t think so much. I think a machine is a machine. I think APMV is a separate entity from me.}'' and 
``\textit{I prioritize efficiency, so I prefer not to have unnecessary communication.}''

In summary, to answer the \textbf{RQ~3}, it may be beneficial to take into account the personality traits of different user groups to better meet their needs and preferences when designing eHMIs.
For instance, simpler and more straightforward interfaces (such as eHMI-T) might be suitable for users who are more introverted and less adventurous, while more complex and diverse interfaces (such as eHMI-AV) could be provided for users who are more extroverted, open, and conscientious.

\subsection{Limitations}
The preference evaluation on eHMIs under different user scenarios may be different.
In this study, we used a simple but common scenario found on campus for testing, which may not fully represent real-life situations, like crowded places with lots of pedestrians.
Furthermore, for some quiet scenes, \eg library and museum, the passengers' user experience of eHMI with voice information cues may be different from the results of this study.

Since one of the objectives of this study was to verify whether the silent eHMI-T could provide a good user experience for a wide range of passengers, we did not balance personality differences between participants before the experiment.
It led to a small number of people choosing eHMI-T and made the standard deviation of neuroticism for passengers who chose eHMI-T too large.
In this regard, we will target introverted passengers in future works to verify if they prefer to use the silent eHMI-T.

Moreover, the participants in this experiment were young Japanese people in their twenties who tend to be more easily receptive to new things than the elderly.
The results may not generalize to other age groups, \eg elderly people, non-student groups or cultural backgrounds.

\section{CONCLUSION}

In this paper, we discussed three eHMI designs, \ie GUI-based eHMI with a text message (eHMI-T), MUI-based eHMI with neutral voice (eHM-NV), and MUI-based eHMI with affective voice (eHMI-AV) used for the APMV-pedestrian communication, from the perspective of APMV passengers.

The field study suggested that the silent GUI-based eHMI-T might not be the most suitable option for APMVs.
This is because when the APMV communicated with a pedestrian, its passenger might encounter difficulties in comprehending the communication, leading to a sense of awkwardness due to the lack of information.
In particular, participants reported the APMV-pedestrian communication was less comprehensive and sufficient, and they felt awkward during the ``silent time", due to the lack of information cues received as passengers.

We also found that the use of voice cues can improve passengers' understanding of the communication between the APMV and pedestrian. 
This enhancement contributes to an overall improved user experience of eHMIs.
Specifically, eHMI-NV has an advantage in pragmatic quality, while ehMI-AV has an advantage in hedonic quality.

This study also highlights that the passengers' personalities should be considered when designing eHMI for APMVs to enhance their user experience.
For instance, simpler and more straightforward interfaces, (such as eHMI-T), might be better suited for passengers with lower levels of extraversion and openness to experience.
Conversely, interfaces with more personality and emotion, (such as eHMI-AV), could be offered to users with higher levels of extraversion, openness to experience, and agreeableness.

In addition, the eHMI-NV is also advantageous for the masses, as the eHMI-NV has also been recognized by the majority of passengers as it balances adequate functionality with acceptable hedonic quality.

\section*{Funding}
This work was supported by JSPS KAKENHI Grant Numbers 20K19846 and 22H00246, Japan.

\section*{Conflict of Interest Statement}

The authors declare that the research was conducted in the absence of any commercial or financial relationships that could be construed as a potential conflict of interest.

\section*{CRediT author statement}

\textbf{Hailong Liu}: Conceptualisation, Methodology, Software, Validation, Formal analysis, Writing - Original Draft, Project administration, Funding acquisition.

\textbf{Yang Li}: Investigation, Methodology, Writing - review \& editing.

\textbf{Zhe Zeng}:  Methodology, Writing - review \& editing.

\textbf{Hao Cheng}: Methodology, Writing - review \& editing.

\textbf{Chen Peng}:  Methodology, Writing - review \& editing.

\textbf{Takahiro Wada}: Methodology, Writing - review \& editing.

\bibliographystyle{cas-model2-names} 
\bibliography{sample.bib}

\begin{thebibliography}{28}
\expandafter\ifx\csname natexlab\endcsname\relax\def\natexlab#1{#1}\fi
\providecommand{\url}[1]{\texttt{#1}}
\providecommand{\href}[2]{#2}
\providecommand{\path}[1]{#1}
\providecommand{\DOIprefix}{doi:}
\providecommand{\ArXivprefix}{arXiv:}
\providecommand{\URLprefix}{URL: }
\providecommand{\Pubmedprefix}{pmid:}
\providecommand{\doi}[1]{\href{http://dx.doi.org/#1}{\path{#1}}}
\providecommand{\Pubmed}[1]{\href{pmid:#1}{\path{#1}}}
\providecommand{\bibinfo}[2]{#2}
\ifx\xfnm\relax \def\xfnm[#1]{\unskip,\space#1}\fi
\bibitem[{Ahn et~al.(2021)Ahn, Lim and Kim}]{ahn2021comparative}
\bibinfo{author}{Ahn, S.}, \bibinfo{author}{Lim, D.}, \bibinfo{author}{Kim,
  B.}, \bibinfo{year}{2021}.
\newblock \bibinfo{title}{Comparative study on differences in user reaction by
  visual and auditory signals for multimodal ehmi design}, in:
  \bibinfo{booktitle}{HCI International 2021-Posters: 23rd HCI International
  Conference, HCII 2021, Virtual Event, July 24--29, 2021, Proceedings, Part
  III 23}, \bibinfo{organization}{Springer}. pp. \bibinfo{pages}{217--223}.
\bibitem[{Ali et~al.(2019)Ali, Lam, Fukuda, Kobayashi and Kuno}]{ali2019smart}
\bibinfo{author}{Ali, S.}, \bibinfo{author}{Lam, A.}, \bibinfo{author}{Fukuda,
  H.}, \bibinfo{author}{Kobayashi, Y.}, \bibinfo{author}{Kuno, Y.},
  \bibinfo{year}{2019}.
\newblock \bibinfo{title}{Smart wheelchair maneuvering among people}, in:
  \bibinfo{booktitle}{International Conference on Intelligent Computing},
  \bibinfo{organization}{Springer}. pp. \bibinfo{pages}{32--42}.
\bibitem[{Bazilinskyy et~al.(2021)Bazilinskyy, Kooijman, Dodou and
  De~Winter}]{bazilinskyy2021should}
\bibinfo{author}{Bazilinskyy, P.}, \bibinfo{author}{Kooijman, L.},
  \bibinfo{author}{Dodou, D.}, \bibinfo{author}{De~Winter, J.},
  \bibinfo{year}{2021}.
\newblock \bibinfo{title}{How should external human-machine interfaces behave?
  examining the effects of colour, position, message, activation distance,
  vehicle yielding, and visual distraction among 1,434 participants}.
\newblock \bibinfo{journal}{Applied ergonomics} \bibinfo{volume}{95},
  \bibinfo{pages}{103450}.
\bibitem[{Brill et~al.(2023)Brill, Payre, Debnath, Horan and
  Birrell}]{brill2023external}
\bibinfo{author}{Brill, S.}, \bibinfo{author}{Payre, W.},
  \bibinfo{author}{Debnath, A.}, \bibinfo{author}{Horan, B.},
  \bibinfo{author}{Birrell, S.}, \bibinfo{year}{2023}.
\newblock \bibinfo{title}{External human--machine interfaces for automated
  vehicles in shared spaces: A review of the human--computer interaction
  literature}.
\newblock \bibinfo{journal}{Sensors} \bibinfo{volume}{23},
  \bibinfo{pages}{4454}.
\bibitem[{Dey et~al.(2021)Dey, van Vastenhoven, Cuijpers, Martens and
  Pfleging}]{dey2021towards}
\bibinfo{author}{Dey, D.}, \bibinfo{author}{van Vastenhoven, A.},
  \bibinfo{author}{Cuijpers, R.H.}, \bibinfo{author}{Martens, M.},
  \bibinfo{author}{Pfleging, B.}, \bibinfo{year}{2021}.
\newblock \bibinfo{title}{Towards scalable ehmis: Designing for av-vru
  communication beyond one pedestrian}, in: \bibinfo{booktitle}{13th
  International Conference on Automotive User Interfaces and Interactive
  Vehicular Applications}, pp. \bibinfo{pages}{274--286}.
\bibitem[{Dou et~al.(2021)Dou, Chen, Tang, Xu and Xue}]{dou2021evaluation}
\bibinfo{author}{Dou, J.}, \bibinfo{author}{Chen, S.}, \bibinfo{author}{Tang,
  Z.}, \bibinfo{author}{Xu, C.}, \bibinfo{author}{Xue, C.},
  \bibinfo{year}{2021}.
\newblock \bibinfo{title}{Evaluation of multimodal external human--machine
  interface for driverless vehicles in virtual reality}.
\newblock \bibinfo{journal}{Symmetry} \bibinfo{volume}{13},
  \bibinfo{pages}{687}.
\bibitem[{Gosling et~al.(2003)Gosling, Rentfrow and Swann~Jr}]{gosling2003very}
\bibinfo{author}{Gosling, S.D.}, \bibinfo{author}{Rentfrow, P.J.},
  \bibinfo{author}{Swann~Jr, W.B.}, \bibinfo{year}{2003}.
\newblock \bibinfo{title}{A very brief measure of the big-five personality
  domains}.
\newblock \bibinfo{journal}{Journal of Research in personality}
  \bibinfo{volume}{37}, \bibinfo{pages}{504--528}.
\bibitem[{Haimerl et~al.(2022)Haimerl, Colley and
  Riener}]{haimerl2022evaluation}
\bibinfo{author}{Haimerl, M.}, \bibinfo{author}{Colley, M.},
  \bibinfo{author}{Riener, A.}, \bibinfo{year}{2022}.
\newblock \bibinfo{title}{Evaluation of common external communication concepts
  of automated vehicles for people with intellectual disabilities}.
\newblock \bibinfo{journal}{Proceedings of the ACM on Human-Computer
  Interaction} \bibinfo{volume}{6}, \bibinfo{pages}{1--19}.
\bibitem[{Hinderks et~al.(2018)Hinderks, Schrepp and
  Thomaschewski}]{hinderks2018benchmark}
\bibinfo{author}{Hinderks, A.}, \bibinfo{author}{Schrepp, M.},
  \bibinfo{author}{Thomaschewski, J.}, \bibinfo{year}{2018}.
\newblock \bibinfo{title}{A benchmark for the short version of the user
  experience questionnaire.}, in: \bibinfo{booktitle}{WEBIST}, pp.
  \bibinfo{pages}{373--377}.
\bibitem[{Kobayashi et~al.(2013)Kobayashi, Suzuki, Sato, Arai, Kuno, Yamazaki
  and Yamazaki}]{Yoshinori2013}
\bibinfo{author}{Kobayashi, Y.}, \bibinfo{author}{Suzuki, R.},
  \bibinfo{author}{Sato, Y.}, \bibinfo{author}{Arai, M.},
  \bibinfo{author}{Kuno, Y.}, \bibinfo{author}{Yamazaki, A.},
  \bibinfo{author}{Yamazaki, K.}, \bibinfo{year}{2013}.
\newblock \bibinfo{title}{Robotic wheelchair easy to move and communicate with
  companions}, in: \bibinfo{booktitle}{CHI '13 Extended Abstracts on Human
  Factors in Computing Systems}, \bibinfo{publisher}{Association for Computing
  Machinery}, \bibinfo{address}{New York, NY, USA}. p.
  \bibinfo{pages}{3079^^e2^^80^^933082}.
\bibitem[{Krei{\ss}ig et~al.(2023)Krei{\ss}ig, Morgenstern and
  Krems}]{kreissig2023blinking}
\bibinfo{author}{Krei{\ss}ig, I.}, \bibinfo{author}{Morgenstern, T.},
  \bibinfo{author}{Krems, J.}, \bibinfo{year}{2023}.
\newblock \bibinfo{title}{Blinking, beeping or just driving? investigating
  different communication concepts for an autonomously parking e-cargo bike
  from a user perspective}.
\newblock \bibinfo{journal}{Human Interaction and Emerging Technologies
  (IHIET-AI 2023): Artificial Intelligence and Future Applications}
  \bibinfo{volume}{70}.
\bibitem[{Lee et~al.(2019)Lee, Sanghavi, Ko and Jeon}]{lee2019autonomous}
\bibinfo{author}{Lee, S.C.}, \bibinfo{author}{Sanghavi, H.},
  \bibinfo{author}{Ko, S.}, \bibinfo{author}{Jeon, M.}, \bibinfo{year}{2019}.
\newblock \bibinfo{title}{Autonomous driving with an agent: Speech style and
  embodiment}, in: \bibinfo{booktitle}{Proceedings of the 11th International
  Conference on Automotive User Interfaces and Interactive Vehicular
  Applications: Adjunct Proceedings}, pp. \bibinfo{pages}{209--214}.
\bibitem[{Li et~al.(2023)Li, Cheng, Zeng, Deml and Liu}]{li2023av}
\bibinfo{author}{Li, Y.}, \bibinfo{author}{Cheng, H.}, \bibinfo{author}{Zeng,
  Z.}, \bibinfo{author}{Deml, B.}, \bibinfo{author}{Liu, H.},
  \bibinfo{year}{2023}.
\newblock \bibinfo{title}{An av-mv negotiation method based on synchronous
  prompt information on a multi-vehicle bottleneck road}.
\newblock \bibinfo{journal}{Transportation Research Interdisciplinary
  Perspectives} \bibinfo{volume}{20}, \bibinfo{pages}{100845}.
\bibitem[{Li et~al.(2021)Li, Cheng, Zeng, Liu and Sester}]{li2021autonomous}
\bibinfo{author}{Li, Y.}, \bibinfo{author}{Cheng, H.}, \bibinfo{author}{Zeng,
  Z.}, \bibinfo{author}{Liu, H.}, \bibinfo{author}{Sester, M.},
  \bibinfo{year}{2021}.
\newblock \bibinfo{title}{Autonomous vehicles drive into shared spaces: ehmi
  design concept focusing on vulnerable road users}, in:
  \bibinfo{booktitle}{2021 IEEE International Intelligent Transportation
  Systems Conference (ITSC)}, \bibinfo{organization}{IEEE}. pp.
  \bibinfo{pages}{1729--1736}.
\bibitem[{Liu et~al.(2020)Liu, Hirayama, Morales and
  Murase}]{liu2020_what_timeing}
\bibinfo{author}{Liu, H.}, \bibinfo{author}{Hirayama, T.},
  \bibinfo{author}{Morales, L.Y.}, \bibinfo{author}{Murase, H.},
  \bibinfo{year}{2020}.
\newblock \bibinfo{title}{What timing for an automated vehicle to make
  pedestrians understand its driving intentions for improving their perception
  of safety?}, in: \bibinfo{booktitle}{2020 IEEE 23rd International Conference
  on Intelligent Transportation Systems (ITSC)}, pp. \bibinfo{pages}{1--6}.
\newblock \DOIprefix\doi{10.1109/ITSC45102.2020.9294696}.
\bibitem[{Liu et~al.(2023)Liu, Hirayama, Saiki and Murase}]{liu2022implicit}
\bibinfo{author}{Liu, H.}, \bibinfo{author}{Hirayama, T.},
  \bibinfo{author}{Saiki, L.Y.M.}, \bibinfo{author}{Murase, H.},
  \bibinfo{year}{2023}.
\newblock \bibinfo{title}{Implicit interaction with an autonomous personal
  mobility vehicle: Relations of pedestrians’ gaze behavior with situation
  awareness and perceived risks}.
\newblock \bibinfo{journal}{International Journal of Human^^e2^^80^^93Computer
  Interaction} \bibinfo{volume}{39}, \bibinfo{pages}{2016--2032}.
\bibitem[{Liu et~al.(2021)Liu, Hirayama and Watanabe}]{liu2021importance}
\bibinfo{author}{Liu, H.}, \bibinfo{author}{Hirayama, T.},
  \bibinfo{author}{Watanabe, M.}, \bibinfo{year}{2021}.
\newblock \bibinfo{title}{Importance of instruction for pedestrian-automated
  driving vehicle interaction with an external human machine interface: Effects
  on pedestrians' situation awareness, trust, perceived risks and decision
  making}, in: \bibinfo{booktitle}{2021 IEEE Intelligent Vehicles Symposium
  (IV)}, \bibinfo{organization}{IEEE}. pp. \bibinfo{pages}{748--754}.
\bibitem[{Oshio et~al.(2012)Oshio, Abe and Cutrone}]{AtsushiOshio2012}
\bibinfo{author}{Oshio, A.}, \bibinfo{author}{Abe, S.},
  \bibinfo{author}{Cutrone, P.}, \bibinfo{year}{2012}.
\newblock \bibinfo{title}{Development, reliability, and validity of the
  japanese version of ten item personality inventory (tipi-j)}.
\newblock \bibinfo{journal}{The Japanese Journal of Personality}
  \bibinfo{volume}{21}, \bibinfo{pages}{40--52}.
\newblock \DOIprefix\doi{10.2132/personality.21.40}.
\bibitem[{Sarsam and Al-Samarraie(2018)}]{sarsam2018towards}
\bibinfo{author}{Sarsam, S.M.}, \bibinfo{author}{Al-Samarraie, H.},
  \bibinfo{year}{2018}.
\newblock \bibinfo{title}{Towards incorporating personality into the design of
  an interface: a method for facilitating users’ interaction with the
  display}.
\newblock \bibinfo{journal}{User Modeling and User-Adapted Interaction}
  \bibinfo{volume}{28}, \bibinfo{pages}{75--96}.
\bibitem[{Schrepp et~al.(2017)Schrepp, Hinderks and
  Thomaschewski}]{schrepp2017design}
\bibinfo{author}{Schrepp, M.}, \bibinfo{author}{Hinderks, A.},
  \bibinfo{author}{Thomaschewski, J.}, \bibinfo{year}{2017}.
\newblock \bibinfo{title}{Design and evaluation of a short version of the user
  experience questionnaire (ueq-s)}.
\newblock \bibinfo{journal}{International Journal of Interactive Multimedia and
  Artificial Intelligence, 4 (6), 103-108.} .
\bibitem[{Sodnik et~al.(2008)Sodnik, Dicke, Toma{\v{z}}i{\v{c}} and
  Billinghurst}]{sodnik2008user}
\bibinfo{author}{Sodnik, J.}, \bibinfo{author}{Dicke, C.},
  \bibinfo{author}{Toma{\v{z}}i{\v{c}}, S.}, \bibinfo{author}{Billinghurst,
  M.}, \bibinfo{year}{2008}.
\newblock \bibinfo{title}{A user study of auditory versus visual interfaces for
  use while driving}.
\newblock \bibinfo{journal}{International journal of human-computer studies}
  \bibinfo{volume}{66}, \bibinfo{pages}{318--332}.
\bibitem[{Tapus and Matari{\'c}(2008)}]{tapus2008user}
\bibinfo{author}{Tapus, A.}, \bibinfo{author}{Matari{\'c}, M.J.},
  \bibinfo{year}{2008}.
\newblock \bibinfo{title}{User personality matching with a hands-off robot for
  post-stroke rehabilitation therapy}, in: \bibinfo{booktitle}{Experimental
  Robotics: The 10th International Symposium on Experimental Robotics},
  \bibinfo{organization}{Springer}. pp. \bibinfo{pages}{165--175}.
\bibitem[{Turk et~al.(1998)Turk, Heimberg, Orsillo, Holt, Gitow, Street,
  Schneier and Liebowitz}]{turk1998investigation}
\bibinfo{author}{Turk, C.L.}, \bibinfo{author}{Heimberg, R.G.},
  \bibinfo{author}{Orsillo, S.M.}, \bibinfo{author}{Holt, C.S.},
  \bibinfo{author}{Gitow, A.}, \bibinfo{author}{Street, L.L.},
  \bibinfo{author}{Schneier, F.R.}, \bibinfo{author}{Liebowitz, M.R.},
  \bibinfo{year}{1998}.
\newblock \bibinfo{title}{An investigation of gender differences in social
  phobia}.
\newblock \bibinfo{journal}{Journal of anxiety disorders} \bibinfo{volume}{12},
  \bibinfo{pages}{209--223}.
\bibitem[{Ullrich(2017)}]{ullrich2017robot}
\bibinfo{author}{Ullrich, D.}, \bibinfo{year}{2017}.
\newblock \bibinfo{title}{Robot personality insights. designing suitable robot
  personalities for different domains}.
\newblock \bibinfo{journal}{i-com} \bibinfo{volume}{16},
  \bibinfo{pages}{57--67}.
\bibitem[{Wang et~al.(2021)Wang, Lee, Kamalesh~Sanghavi, Eskew, Zhou and
  Jeon}]{wang2021vehicle}
\bibinfo{author}{Wang, M.}, \bibinfo{author}{Lee, S.C.},
  \bibinfo{author}{Kamalesh~Sanghavi, H.}, \bibinfo{author}{Eskew, M.},
  \bibinfo{author}{Zhou, B.}, \bibinfo{author}{Jeon, M.}, \bibinfo{year}{2021}.
\newblock \bibinfo{title}{In-vehicle intelligent agents in fully autonomous
  driving: The effects of speech style and embodiment together and separately},
  in: \bibinfo{booktitle}{13th International Conference on Automotive User
  Interfaces and Interactive Vehicular Applications}, pp.
  \bibinfo{pages}{247--254}.
\bibitem[{Watanabe et~al.(2015)Watanabe, Ikeda, Morales, Shinozawa, Miyashita
  and Hagita}]{watanabe2015communicating}
\bibinfo{author}{Watanabe, A.}, \bibinfo{author}{Ikeda, T.},
  \bibinfo{author}{Morales, Y.}, \bibinfo{author}{Shinozawa, K.},
  \bibinfo{author}{Miyashita, T.}, \bibinfo{author}{Hagita, N.},
  \bibinfo{year}{2015}.
\newblock \bibinfo{title}{Communicating robotic navigational intentions}, in:
  \bibinfo{booktitle}{2015 IEEE/RSJ International Conference on Intelligent
  Robots and Systems (IROS)}, \bibinfo{organization}{IEEE}. pp.
  \bibinfo{pages}{5763--5769}.
\bibitem[{Zentner et~al.(2023)Zentner, Lee, Dueck and
  Masuda}]{zentner2023cultural}
\bibinfo{author}{Zentner, K.E.}, \bibinfo{author}{Lee, H.},
  \bibinfo{author}{Dueck, B.S.}, \bibinfo{author}{Masuda, T.},
  \bibinfo{year}{2023}.
\newblock \bibinfo{title}{Cultural and gender differences in social anxiety:
  The mediating role of self-construals and gender role identification}.
\newblock \bibinfo{journal}{Current Psychology} \bibinfo{volume}{42},
  \bibinfo{pages}{21363--21374}.
\bibitem[{Zhang et~al.(2022)Zhang, Barbareschi, Ramirez~Herrera, Carlson and
  Holloway}]{zhang2022understanding}
\bibinfo{author}{Zhang, B.}, \bibinfo{author}{Barbareschi, G.},
  \bibinfo{author}{Ramirez~Herrera, R.}, \bibinfo{author}{Carlson, T.},
  \bibinfo{author}{Holloway, C.}, \bibinfo{year}{2022}.
\newblock \bibinfo{title}{Understanding interactions for smart wheelchair
  navigation in crowds}, in: \bibinfo{booktitle}{Proceedings of the 2022 CHI
  Conference on Human Factors in Computing Systems}, pp.
  \bibinfo{pages}{1--16}.

\end{thebibliography}

\end{document}